\documentclass[aps,prl,twocolumn,superscriptaddress]{revtex4-1}
\usepackage[utf8]{inputenc}
\usepackage[english]{babel}
\usepackage{amsmath}
\usepackage{amsfonts}
\usepackage{amssymb}
\usepackage{graphicx}
\usepackage{hyperref}
\usepackage{physics}

\usepackage{braket}
\usepackage[squaren,thickqspace]{SIunits}

\usepackage[usenames]{color}
\usepackage{soul}

\newcommand{\fR}{\ensuremath{f_{\text{r}}}}
\newcommand{\fSR}{\ensuremath{f_{\text{SR}}}}
\newcommand{\geff}{\ensuremath{g_\text{eff}}}
\newcommand{\BMR}{\ensuremath{B_\text{MR}}}

\newcommand{\Teff}{\ensuremath{T_\text{eff}}}

\newcommand{\JA}{\ensuremath{J^\mathrm{A}}}
\newcommand{\kB}{\ensuremath{k_\mathrm{B}}}
\newcommand{\muB}{\ensuremath{\mu_\mathrm{B}}}

\begin{document}

\title{Strong Coupling of Microwave Photons to Antiferromagnetic Fluctuations in an Organic Magnet}

\author{Matthias Mergenthaler}
\email{matthias.mergenthaler@materials.ox.ac.uk}
\affiliation{Department of Materials, University of Oxford, Oxford OX1 3PH, United Kingdom}
\affiliation{Clarendon Laboratory, Department of Physics, University of Oxford, Oxford OX1 3PU, United Kingdom}

\author{Junjie Liu}
\affiliation{Department of Materials, University of Oxford, Oxford OX1 3PH, United Kingdom}

\author{Jennifer J. \surname{Le Roy}}
\affiliation{Department of Materials, University of Oxford, Oxford OX1 3PH, United Kingdom}

\author{Natalia Ares}
\affiliation{Department of Materials, University of Oxford, Oxford OX1 3PH, United Kingdom}

\author{Amber L. Thompson}
\affiliation{Chemical Crystallography, Chemistry Research Laboratory, University of Oxford, Oxford OX1 3TA, United Kingdom}

\author{Lapo Bogani}
\affiliation{Department of Materials, University of Oxford, Oxford OX1 3PH, United Kingdom}

\author{Fernando Luis}
\affiliation{Instituto de Ciencia de Materiales de Aragón (CSIC-U. de Zaragoza), 50009 Zaragoza, Spain}

\author{Stephen J. Blundell}
\affiliation{Clarendon Laboratory, Department of Physics, University of Oxford, Oxford OX1 3PU, United Kingdom}

\author{Tom Lancaster}
\affiliation{Durham University, Centre for Materials Physics, Department of Physics, Durham DH1 3LE, United Kingdom}

\author{Arzhang Ardavan}
\affiliation{Clarendon Laboratory, Department of Physics, University of Oxford, Oxford OX1 3PU, United Kingdom}

\author{G. Andrew D. Briggs}
\affiliation{Department of Materials, University of Oxford, Oxford OX1 3PH, United Kingdom}

\author{Peter J. Leek}
\affiliation{Clarendon Laboratory, Department of Physics, University of Oxford, Oxford OX1 3PU, United Kingdom}

\author{Edward A. Laird}
\email{edward.laird@materials.ox.ac.uk}
\affiliation{Department of Materials, University of Oxford, Oxford OX1 3PH, United Kingdom}

\date{\today}

\begin{abstract}
Coupling between a crystal of di(phenyl)-(2,4,6-trinitrophenyl)iminoazanium radicals and a superconducting microwave resonator is investigated in a circuit quantum electrodynamics (circuit QED) architecture. The crystal exhibits paramagnetic behavior above 4~K, with  antiferromagnetic correlations appearing below this temperature, and we demonstrate strong coupling at base temperature. The magnetic resonance acquires a field angle dependence as the crystal is cooled down, indicating anisotropy of the exchange interactions.
These results show that multispin modes in organic crystals are suitable for circuit QED, offering a platform for their coherent manipulation.
They also utilize the circuit QED architecture as a way to probe spin correlations at low temperature.
\end{abstract}
\maketitle

Hybrid circuit quantum electrodynamics (circuit QED) using spin ensembles coupled to microwave resonators~\hbox{\cite{Wesenberg2009, Xiang2013,Schuster2010a,Kubo2010,Amsuss2011,Ranjan2013,Clauss2013}} has potential use in quantum memories~\hbox{\cite{Kubo2011,Grezes2015}} as well as for microwave-to-optical conversion~\hbox{\cite{Blum2015}}. The first demonstrations used paramagnetic ensembles, but correlated states such as ferrimagnets lead to stronger coupling because of their high spin density~\hbox{\cite{Huebl2013, Cao2015}}. However, this comes at the price of on-chip magnetic fields, to which both superconducting qubits (used as processors) and SQUID arrays (used for cavity tuning) are sensitive.
Antiferromagnetic spin ensembles circumvent this obstacle by combining high spin densities with no net magnetization. Perpendicular spin axis alignments of antiferromagnetic domains can also be used as a classical memory, which is robust against high magnetic fields, invisible to magnetic sensors, and can be packed with high density. Antiferromagnetic memory devices can be manipulated and read out via electrical currents~\hbox{\cite{Marti2014,Jungwirth2016}}.
Furthermore, antiferromagnetic heterostructures would combine spintronic and magnonic functionalities~\hbox{\cite{Troncoso2015}}.
Harnessing these possibilities makes it necessary to understand the range of interactions that occur in antiferromagnetic systems.
As model systems, organic magnets can be engineered chemically to create well-defined magnetic interactions~\hbox{\cite{GoddardPRL2012,LiuIC2016}}, which could be probed via circuit QED to test models of magnetism in different dimensions~\hbox{\cite{Kahn1993a}}.
Characteristic interaction strengths in organic magnets are such that these materials typically approach or undergo a phase transition only at mK temperatures~\hbox{\cite{Chiarelli1993, Sichelschmidt2003, Blundell2004, Wiemann2015}}, making them difficult to study with conventional electron spin resonance (ESR).
Strong coupling to antiferromagnetic correlations, which has not yet been achieved in the circuit QED architecture, would allow these materials to be studied at low temperatures, low microwave frequencies, and low magnetic fields.

Here we demonstrate strong coupling between microwave modes of a superconducting resonator and a crystallized organic radical, di(phenyl)-(2,4,6-trinitrophenyl)iminoazanium (DPPH).
In this material antiferromagnetic correlations become evident in spin resonance at a temperature $T\sim 4$~K and below, although no magnetic ordering is observed down to a temperature of 16~mK~\footnote{See Supplemental Material at URL, which includes Refs.~\cite{Altomare1994,Parois2015,Betteridge2003,Cooper2010,Spek2003,Blundell2001,Bellido2013,Tosi2014,Jenkins2013}, for resonator characterization, crystal characterization, susceptibility measurements, muon spectroscopy, data set of Crystal II and calculations of the temperature-dependent frequency shift, the single spin coupling and number of radicals in the crystal.}. We measure coupling both to spin excitations (in the paramagnetic phase at high temperature $T\gtrsim 4$~K), and to excitations showing antiferromagnetic correlations at lower temperature~\cite{Prokhorov1963,TeitelBaum1981}.
By studying the angle dependence of the magnetic resonance, we investigate the anisotropy of the exchange interactions, evident from a separation of parallel and perpendicular resonances as the crystal is cooled.
We measure the ensemble coupling as a function of temperature, which shows paramagnetic behavior above $T\sim 500$~mK but becomes temperature independent below $T\sim 50$~mK. The spin modes deviate from paramagnetic behavior due to antiferromagnetic (AFM) fluctuations being present, despite being above the AFM phase transition temperature.

\begin{figure}[h!]
\includegraphics[width=0.95\columnwidth]{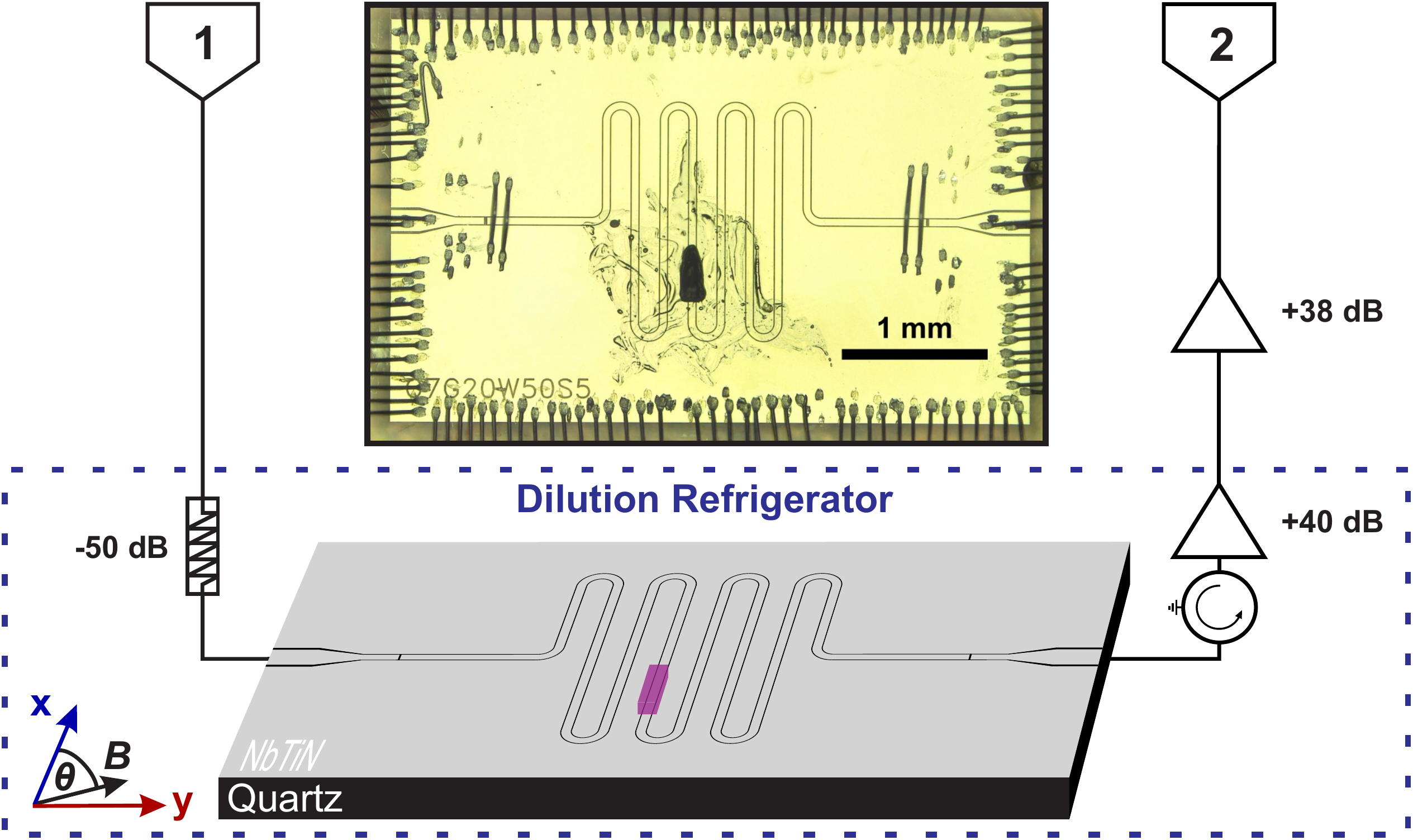}
\caption{\label{fig1}Experimental schematic. The coplanar resonator (inset photograph) is mounted in a dilution refrigerator and measured via two-port microwave transmission. A DPPH crystal (purple in schematic, black in photograph) is attached with vacuum grease near the magnetic field antinode of the resonator's fundamental mode. Axes of the in-plane static magnetic field are indicated.}
\end{figure}

To fabricate the superconducting resonator, a $\unit{110}{\nano\meter}$ NbTiN film was sputtered onto a quartz substrate, and  patterned using optical lithography and reactive ion etching. The measured resonator (Fig.~\ref{fig1}) has a signal line width of $w=\unit{50}{\micro\metre}$ and a separation of $s=\unit{5.3}{\micro\metre}$ from the lateral ground planes for $\unit{50}{\ohm}$ impedance matching.
Single crystals were grown via a saturated solution of DPPH in toluene, sitting in a hexane bath at 5~$^\circ$C over two weeks. Using this method DPPH crystallizes in a triclinic P-1 space group with a unit cell consisting of four DPPH, one hexane, and one toluene molecule~\cite{Note1}.
The largest crystals from two identically prepared growth batches were measured; results from one crystal (crystal~I) are presented here, while results from crystal~II, with similar behavior, are shown in the Supplemental Material~\cite{Note1}.
Each measured crystal was attached near the magnetic field antinode of the cavity fundamental mode, with the long axis aligned along the CPWR, defining the $x$ axis.
Measurements were performed in a dilution refrigerator in an in-plane magnetic field $\mathbf{B} \equiv(B_x, B_y, 0)$.

The device was measured by transmission spectroscopy using a microwave network analyzer.
In zero magnetic field and at $T=\unit{15}{\milli\kelvin}$, the resonator (with crystal attached) exhibits a fundamental mode at frequency $\omega_0/2\pi=f_0=\unit{5.92}{\giga\hertz}$ and a loaded quality factor of $Q_{\textrm{L}}=1.51\times 10^{4}$~\cite{Note1}.
An external magnetic field of magnitude $B\equiv |\mathbf{B}|=\unit{165}{\milli\tesla}$ applied along $x$ (along~$y$) reduces this to $Q_{\textrm{L}}=1.17\times 10^{4}$ ($Q_{\textrm{L}}=1.04\times 10^4$).

\begin{figure}[t!]
\centering
\includegraphics[width=\columnwidth]{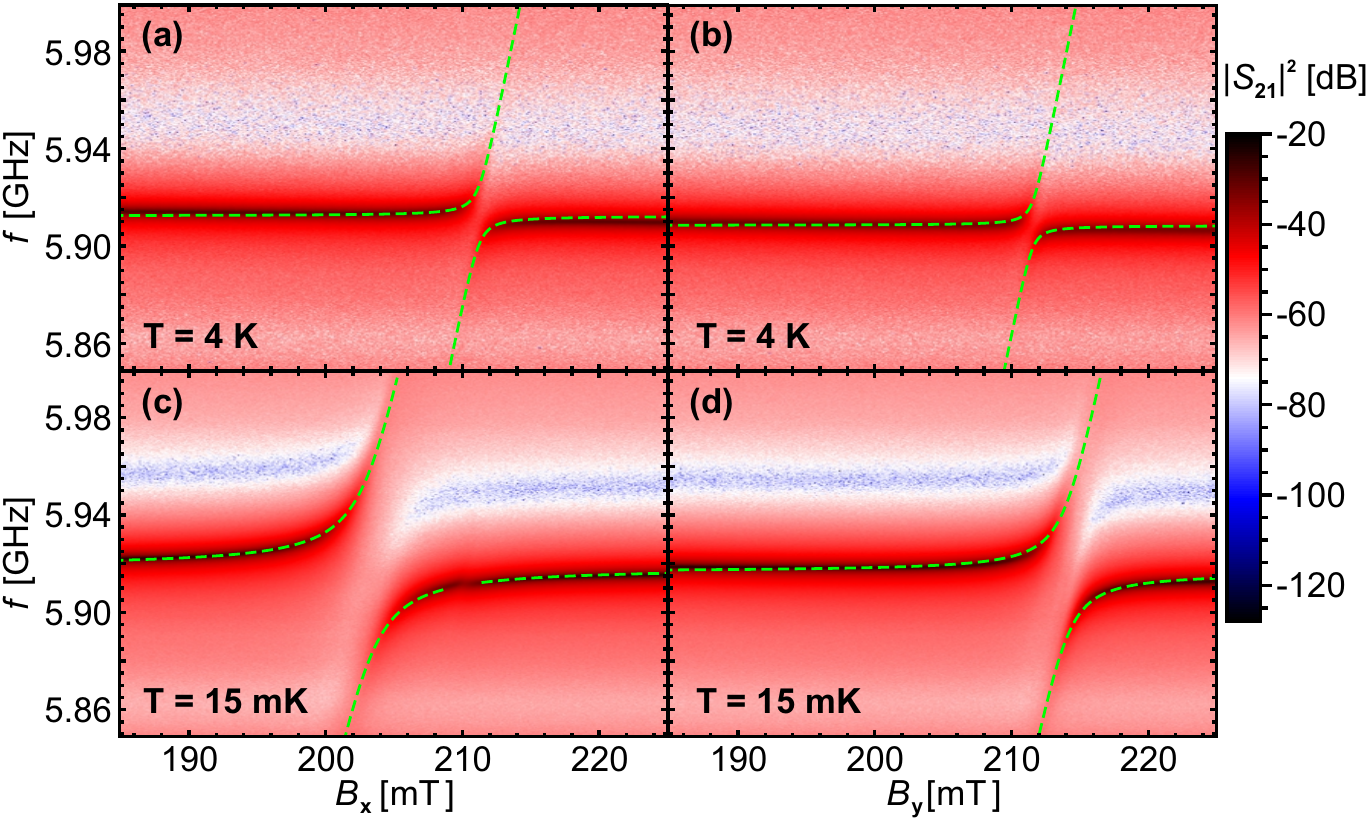}
\caption{\label{fig2}Transmission as a function of external magnetic field $B_{x,y}$ and resonator probe frequency $f$, measured at two different temperatures. Transmission maxima occur at resonance frequencies of the combined system, with anticrossings indicating hybridization between crystal magnetic resonances and the cavity modes. Superimposed on each panel are fits to the resonance frequencies (dashed lines) using Eq.~(\ref{eq:anticrossing}).
}
\end{figure}

To probe coupling to the crystal, the resonator transmission $\left|S_{21}\right|^2$ is measured at two different temperatures as a function of frequency $f$ and magnetic field (Fig.~\ref{fig2}). The bare cavity mode is evident as a transmission peak that is nearly field independent. As the magnetic field is swept, the spin resonance frequency $\fSR$ is tuned through degeneracy with the cavity frequency $\omega_{\text{r}}/2\pi=\fR$, giving rise to an anticrossing when $\fSR \approx \fR$.

Because of the large number of molecular spins, it is appropriate to parametrize the coupling to the resonator by an effective ensemble coupling $\geff$~\cite{Wesenberg2009,Schuster2010,Kubo2010}.
To extract $g_{\textrm{eff}}$, the system is modeled as two coupled oscillators, giving for the hybridized resonance frequency~\cite{Abe2011}
\begin{equation}
\omega_\pm=\omega_{\text{r}}+\frac{\Delta}{2}\pm\frac{1}{2}\sqrt{\Delta^2+4g_{\textrm{eff}}^2},
\label{eq:anticrossing}
\end{equation}
where $\omega_\pm/2\pi=f_\pm$, $\Delta=g\mu_{\textrm{B}}\left(B_{x,y}-B_{\textrm{MR}}\right)/\hbar$ is the frequency detuning and $\BMR$ is the magnetic resonance (MR) field.
Fitting the transmission peak locations in Fig.~2 to Eq.~(\ref{eq:anticrossing}) and assuming a fixed Land\'e factor $g=2.0037$~\cite{Ghirri2015b} gives the fit parameters $\geff$ and $\BMR$ shown in Table~\ref{tab:param} for the two field directions and temperatures.

The spin dephasing rate $\gamma(T)$ is deduced by fitting a standard input-output model \cite{Abe2011,Clerk2010,Schuster2010,Huebl2013,Zollitsch2015}
\begin{equation}\label{eqn:S21}
\left|S_{21}(\omega)\right|^2=\left|\frac{\kappa_c}{i(\omega-\omega_{\text{r}})-\kappa+\frac{g_{\textrm{eff}}^2}{i(\omega-\omega_\mathrm{MR})-\gamma}}\right|^2,
\end{equation}
where  $\kappa_c$ is the coupling rate to the external microwave circuit and $2\kappa/2\pi\equiv f_0/Q_{\textrm{L}}$ is the total relaxation rate of the resonator. We use Eq.~(\ref{eqn:S21}) to fit $\left|S_{21}(\omega)\right|^2$ at the resonance fields $\BMR$, taking $\kappa_c$ and $\gamma$ as fit parameters and holding constant the parameters $\geff$, $\omega_\mathrm{r}$ and $\kappa$ deduced above. Extracted values of $\gamma$ are shown in Table~\ref{tab:param}.

	\begin{table}[b!]
		\centering
		\setlength\tabcolsep{2pt}
		\begin{tabular}{cccccc}
			\hline \hline
			$T$ (K) & Axis	& $B_{\textrm{MR}}$~(mT) 	& $g_{\textrm{eff}}/2\pi$~(MHz) 	& $\gamma/2\pi$ (MHz)	& $C$ \\
			\hline
			4		& $x$ 	& $211.19\pm 0.05$ 			& $12.1 \pm 0.4$			&  $15.0 \pm 0.2$ 	& 18 \\
			4		& $y$ 	& $211.53\pm 0.05$ 			& $\phantom{1}9.6\pm 0.3$ 	&  $15.0 \pm 0.2$ 	& 10 \\
			0.015 	& $x$ 	& $203.12\pm 0.02$ 			& $38.7\pm 0.1$ 			& $29.6\pm 0.2$ 	& 200 \\
			0.015 	& $y$ 	& $213.75\pm 0.05$ 			& $26.9\pm 0.3$ 			&  $25.5\pm 0.4$ 	& 102 \\
			\hline
			\hline
		\end{tabular}
		\caption{Resonance parameters extracted from Fig.~\ref{fig2} for different temperatures and magnetic field orientations.}
		\label{tab:param}
	\end{table}

A dimensionless measure of the coupling efficiency is the cooperativity $C\equiv \geff^2/\kappa\gamma$.
We extract this parameter for each temperature and field axis (Table~\ref{tab:param}). Already at $T=4$~K,  the system is in the regime of high cooperativity ($C>1$), implying coherent transfer of excitations from the microwave field to the ensemble, while at $T=15$~mK the strong coupling condition $g_{\textrm{eff}}\gg\kappa,\gamma$ is reached for $\mathbf{B}$ along $x$, where the ensemble coupling is faster than the decay of both the spin ensemble and the cavity.

\begin{figure}[tb]
\centering
\includegraphics[width=\columnwidth]{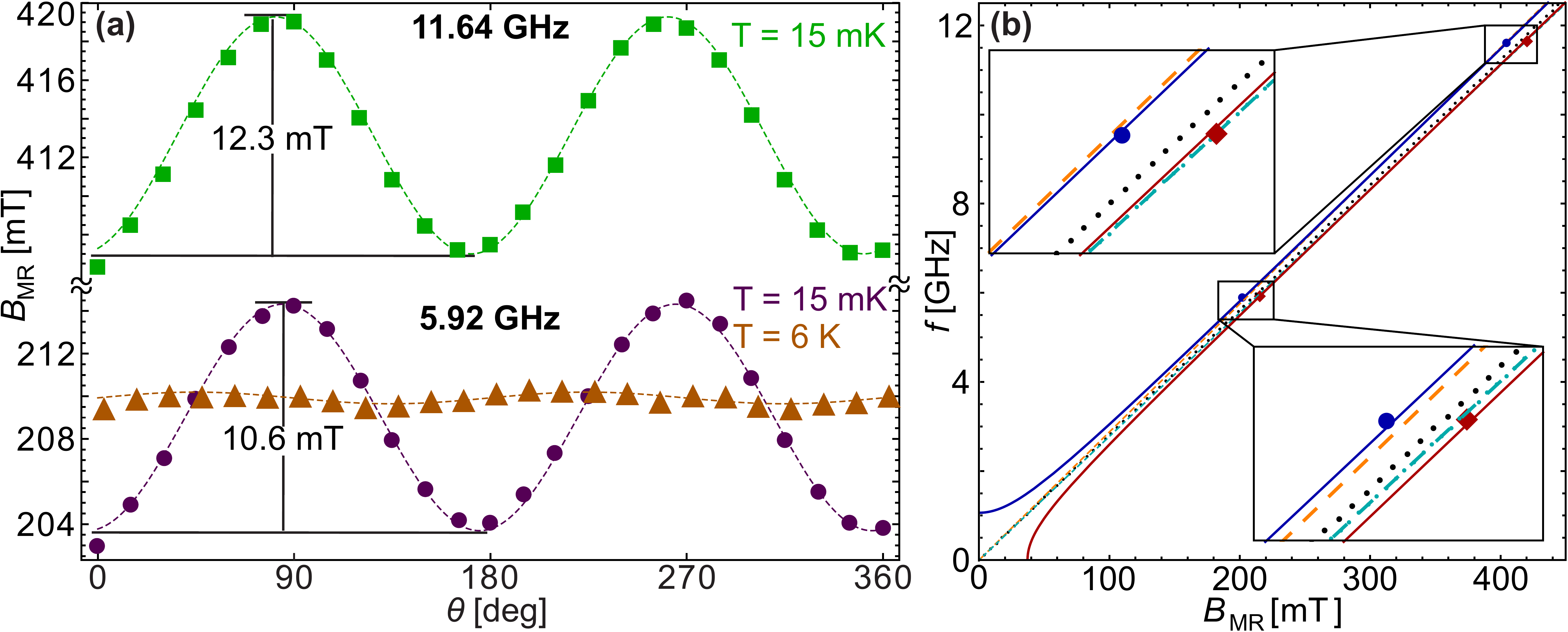}
\caption{\label{fig3} (a) Resonance magnetic field as a function of field angle $\theta$. Measuring at $T=15$~mK, the resonance field varies sinusoidally with $\theta$, with amplitude $\Delta B_0=10.6$~mT for the fundamental mode ($f_0 = 5.92$~GHz, circles) and $\Delta B_1=12.3$~mT for the first harmonic mode ($f_1=11.64$~GHz, squares). At high temperature, the fundamental mode shows nearly isotropic resonance (triangles). (b) Plot of the MR frequency as a function of resonance magnetic field. Data points are the resonance magnetic fields along  $x$ (circles) and along $y$ (diamonds), taken from the maximum and minimum data points of (a) for data at the fundamental or first harmonic mode. The black dotted line is the PM dispersion relation with Land\'e factor $g=2.0037$. The dashed orange and dot-dashed cyan lines are fits using a PM dispersion relation, with separate $g$ factors along the two axes taken as fit parameters. From the insets it is apparent that these fits do not describe the data well. Red and blue solid curves are a fit to the AFM dispersion relation in Eq.~(\ref{eqn:AFM_model_1}), which agrees well with the data.}
\end{figure}

We now show that the crystal exhibits antiferromagnetic correlations at low temperature. Whereas at high temperature [Figs.~\ref{fig2}(a) and \ref{fig2}(b)], the anticrossing field $\BMR$ is nearly independent of angle, at $T=15$~mK there is a pronounced anisotropy~[Figs.~\ref{fig2}(c) and \ref{fig2}(d)]. This is explored further in Fig.~\ref{fig3}(a), which compares the dependence of $\BMR$ on field angle $\theta$ at $T=6$~K and $T=15$~mK. Measuring near the fundamental cavity mode $f_0$, the angle dependence is well fit by
$B_\mathrm{MR}=B_\mathrm{MR}^\mathrm{offset} + \Delta B_i \sin^2 (\theta + \Delta \theta)$, with offsets $B_\mathrm{MR}^\mathrm{offset}$ and $\Delta \theta$ together with anisotropy $\Delta B_i$ as fit parameters, where $i\in\{0,1\}$ labels the cavity mode.  At low temperature, we find $\Delta B_0 = 10.6$~mT, whereas at 6~K there is almost no angle dependence.

At high temperature, this is consistent with a paramagnet with nearly isotropic $g$ factor \cite{Note1}. Anisotropy at lower temperature could arise from field screening by the superconductor, from temperature-dependent $g$-factor anisotropy, from trapped flux in the magnet coils, or from a transition to magnetic correlations in the crystal. Field screening is excluded by measurements with different crystal orientation~\cite{Note1}. To exclude $g$-factor anisotropy, we repeated the measurement at the first harmonic of the resonator [$f_1=11.64$~GHz, upper trace in Fig.~\ref{fig3}(b)]. Whereas $g$-factor anisotropy would lead to $\Delta B_1 = 2 \Delta B_0 $, in fact we find $\Delta B_1=12.3$~mT~$\approx \Delta B_0$. Trapped flux in the coils is also excluded by the temperature dependence, since the coils are thermally isolated from the sample. We therefore deduce an onset of AFM correlations between 15~mK and 4~K.

To confirm antiferromagnetic behavior, we plot the magnetic resonance dispersion relation for the two principal axes [Fig.~\ref{fig3}(b)]. Although each branch contains only two data points, they clearly do not satisfy a paramagnetic (PM) dispersion relation $f = g\muB \BMR/h$ (dotted/dashed/dot-dashed lines on figure), even allowing for $g$-tensor anistropy. However, they are well fit by an AFM dispersion relation~\cite{Katsumata2000} derived from a two-sublattice model with a molecular-field approximation at zero temperature~\cite{Nagamiya1955}:
\begin{equation}\label{eqn:AFM_model_1}
f=\frac{g\mu_{\textrm{B}}}{h}\sqrt{\BMR^2\pm K},
\end{equation}
with the + (-) branches describing field alignment parallel (perpendicular)  to the anisotropy axis. Here the fit parameters are $K$, which parametrizes the exchange and anisotropy field of the crystal and separate $g$ factors $g_x$ and $g_y$ for the two field directions~\cite{Magarino1978, Katsumata2000}. Fitting all four data points simultaneously, the best fit parameters are $K=0.0014$~mT$^2$, $g_x=2.04$, and $g_y=1.99$, similar to a previously reported value $g=2.0037$ in the PM phase~\cite{Ghirri2015b}.
At low temperature, the magnetic resonance excitations are no longer single spin flips, but antiferromagnetic fluctuations.

\begin{figure}[tb]
\centering
\includegraphics[width=\columnwidth]{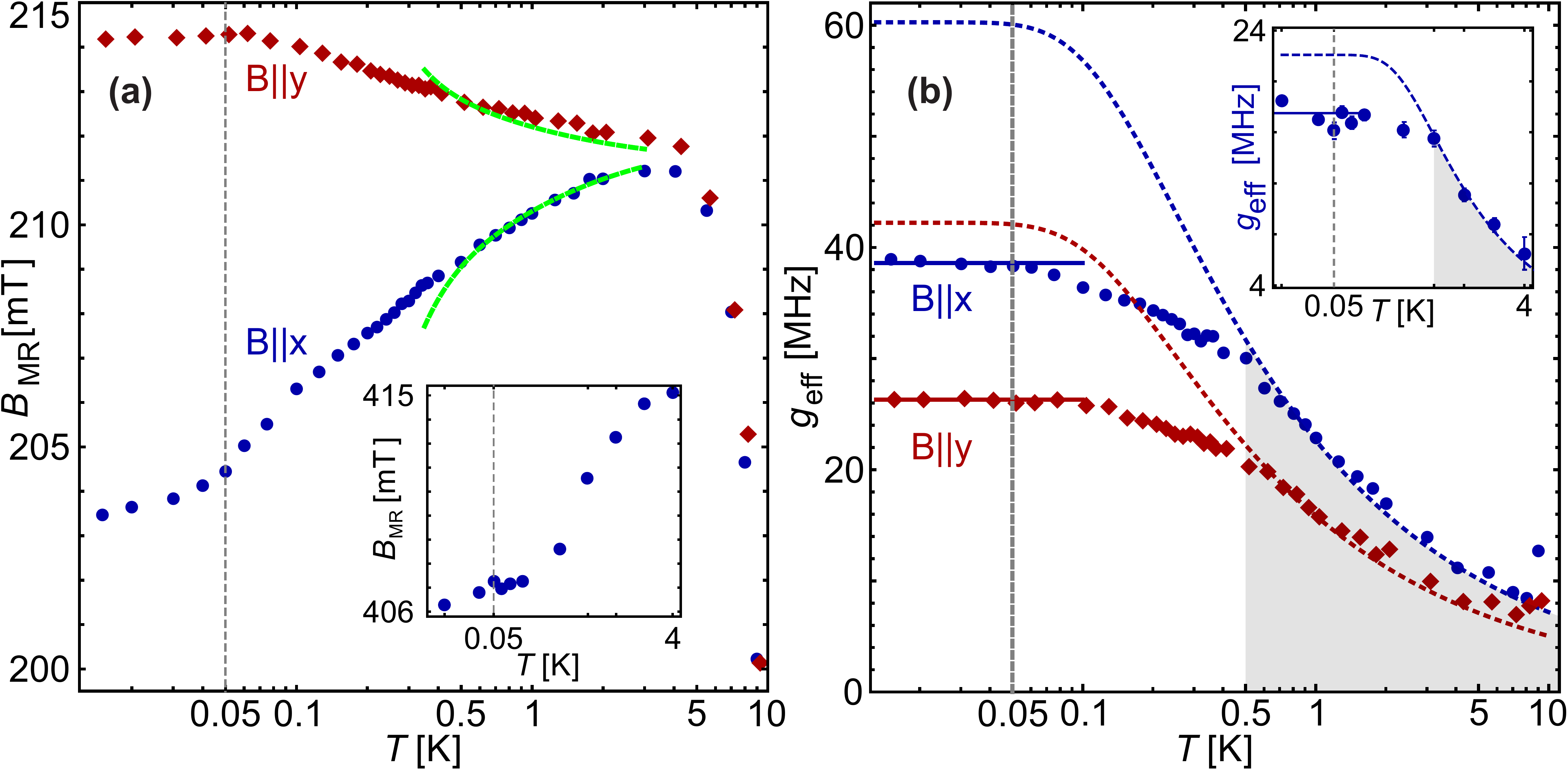}
\caption{\label{fig4} (a) Temperature evolution of $g_{\textrm{eff}}$ for $\mathbf{B}$ applied along $x$ and $y$. Above $T\sim 500$~mK, the data agree with a PM model [dashed lines, fit to Eq.~(\ref{eq:geff}) over the shaded temperature range]. Inset right: similar data and fits at the resonator's first harmonic. Inset left: effective spin-temperature calculated with Eq.~(\hbox{\ref{eq:geff}}) (points). (b) Filled symbols: resonance magnetic field along $x$ and $y$ as a function of temperature. As temperature decreases, the resonance magnetic field moves away from its paramagnetic value (assuming $g=2.0037$). At intermediate temperatures, both branches are fit by a spin chain model (solid curves; see text). Unfilled symbols: $\BMR$ along $x$ and $y$ as a function of effective temperature according to Eq.~(\hbox{\ref{eq:geff}}). The data for $T\leq 3$~K is fit by a spin chain model (dashed curve; see text). Inset: similar data at the first harmonic mode.}
\end{figure}

The temperature evolution of the effective polarization can be  studied via the coupling strength $g_{\textrm{eff}}$  [Fig.~\ref{fig4}(a)]. Above 50~mK, $g_{\textrm{eff}}$ decreases with increasing temperature, as expected from thermal depolarization of the spin ensemble.
For a paramagnet, the effective coupling is~\cite{Wesenberg2009}
\begin{equation}
g_{\textrm{eff}}(T)=g_s\sqrt{N_{\textrm{P}}(T)} = g_s\sqrt{N\,\textrm{tanh}\left(hf/2k_{\textrm{B}}T\right)},
\label{eq:geff}
\end{equation}
where $g_s$ is the root-mean-square coupling per individual spin and $N_{\textrm{P}}(T)$ is the net number of polarized spins out of $N$ coupled radicals. Above $T=0.5$~K [shaded region of Fig.~\ref{fig4}(a)], Eq.~\eqref{eq:geff} gives a good fit to the data; calculating $g_s/2\pi=5$~Hz from the geometry of the resonator and taking the number of coupled radicals as a fit parameter gives $N_x=1.5\times 10^{14}$ for $\mathbf{B}$ along $x$. This is in fair agreement with $N=1.7\times 10^{14}$ estimated from the geometry of the crystal. The data for $\mathbf{B}$ along $y$ give a smaller value $N_y=7.1\times 10^{13}$, as expected from the smaller perpendicular overlap with the alternating cavity field.
High cooperativity ($C>1$) is already reached far above base temperature, for example at $T=0.5$~K, where $C_x=66$ and $C_y=28$.
The agreement with the two-level model [Eq.~(\ref{eq:geff})] confirms that the magnetic resonance spectroscopy probes a transition from the spin ground state (rather than between two excited states).

Below 0.5~K, $g_\mathrm{eff}$ is found to be smaller than the fits would predict. This may reflect screening of each spin by its  neighbors as the antiferromagnetic phase is approached (although the $\sqrt{N_\mathrm{P}}$ enhancement of $\geff$ is still expected to apply~\cite{Cao2015}). It may also reflect a failure of the spin ensemble to thermalize. By comparing the measured $\geff(T)$ with the value predicted by Eq.~\eqref{eq:geff}, an effective spin temperature $\Teff$ can be extracted~[Fig.~\ref{fig4}(a) inset left].
At the lowest temperature, the effective number of coupled spins  is $N_\mathrm{P} = (\geff/g_s)^2 \approx 5.9\times 10^{13}$ for $\mathbf{B}$ along $x$.
Similar behavior is observed at the resonator's first harmonic mode [Fig.~\ref{fig4}(a) inset right], with smaller overall coupling because the crystal is not located at a field antinode.

We now study the temperature dependence of the magnetic resonance, which gives experimental insight into the spin correlations, where analytical solutions for models of interacting spins in three dimensions do not exist. The shift of the magnetic resonance frequency away from the high-temperature (paramagnetic) value is a measure of short-range correlations.
Filled symbols in Fig.~\ref{fig4}(b) show the magnetic resonance field as a function of cryostat temperature for parallel and perpendicular field alignment.
Both data sets exhibit a kink at $T\sim\unit{50}{\milli\kelvin}$ which could suggest a phase transition, and indeed such a transition to an AFM state at $T\sim 0.3$~K has been previously observed in DPPH~\cite{Prokhorov1963,TeitelBaum1981}. However, in our sample, separate investigations using ac susceptibility and muon spectroscopy~\cite{Note1} show that there is no phase transition down to $T=16$~mK.
The transition temperature in DPPH is known to vary widely depending on the crystallizing solvent~\cite{Kessel1973}, and the incorporated toluene and hexane in our crystal presumably inhibits ordering at accessible temperatures~\cite{Note1}.
For this reason, we attribute the low-temperature kink in Fig.~\ref{fig4}(b) (filled symbols) to the failure of the spins to thermalize inside the resonator.
This interpretation is supported by plotting the same data as a function of the spin temperature $T_{\textrm{eff}}$ [extracted as in Fig.~\hbox{\ref{fig4}}(a) left inset], which shows that the kink disappears [Fig.~\hbox{\ref{fig4}}(b), unfilled symbols].
At high temperature ($T\gtrsim5$~K), the resonances shift to lower field because of the (independently measured) decrease in cavity  frequency due to kinetic inductance.

The temperature dependence of the resonance frequencies is simulated by calculating the short range spin-spin correlations between DPPH molecules. The spin Hamiltonian is
\begin{equation}
\mathcal{H} = \mathcal{H}^0 + \mathcal{H}^\prime,
\label{eq:Hamiltonian}
\end{equation}
where $\mathcal{H}^0 = -2\sum_{i,j} J_{ij}\mathbf{S}_i\cdot\mathbf{S}_j - g\mu_\mathrm{B}\sum_{i} \mathbf{B}\cdot\mathbf{S}_i$ incorporates isotropic exchange and Zeeman energy, and $\mathcal{H}^\prime$ represents the anisotropic exchange between molecules, e.g. dipole-dipole interactions.
Here $\mathbf{S}_i = \{ S_i^x, S_i^y, S_i^z \}$ is the spin of the $i\mathrm{th}$ molecule, and $J_{ij}<0$ is the isotropic exchange. Equation~(\ref{eq:Hamiltonian}) assumes an isotropic $g$ tensor, which is not required by symmetry but is justified experimentally by the isotropy of the magnetic resonance field well above the phase transition [Fig.~\ref{fig3}(a)]. We neglect the bulk permeability of the material. In the absence of anisotropy ($\mathcal{H}^\prime = 0$), Eq.~(\ref{eq:Hamiltonian}) leads to a temperature independent ESR resonance frequency with $f = g\mu_\mathrm{B}B$, which is identical to the ESR resonance for noninteracting spins, despite the isotropic interaction~\cite{Oshikawa2002}. Any shift of this resonance frequency indicates an effect of $\mathcal{H}^\prime$. Assuming $\mathcal{H}^0 \gg \mathcal{H}^\prime$, the frequency shift is~\cite{Nagata1972,Oshikawa2002,Maeda2005}
\begin{equation}
\label{eqn:tmodel}
h\delta f = -\frac{\langle [[\mathcal{H}^\prime,S^+],S^-]\rangle}{2\langle S^z \rangle},
\end{equation}
where $\langle ... \rangle$ indicates the temperature-dependent expectation value, $\mathbf{S}\equiv\sum_i \mathbf{S}_i$ is the total spin operator, and $S^\pm\equiv S^x \pm iS^y$.

To gain insight into the role of antiferromagnetic fluctuations, we employ a simple model of a one-dimensional uniaxial anisotropic antiferromagnet~\cite{Nagata1972}. This is also suggested by the crystal packing, where solvent molecules may act as blocks between chains~\cite{Note1}. We therefore have $\mathcal{H}^0 = -2J\sum_i\mathbf{S}_i\cdot\mathbf{S}_{i+1} - g\mu_\mathrm{B}\sum_i{\mathbf{B}\cdot\mathbf{S}_i}$ and $\mathcal{H}^\prime = 2J^\mathrm{A} \sum_i S^x_i S^x_{i+1}$. In a classical approximation, expected to be valid at high temperature, the frequency shift Eq.~(\ref{eqn:tmodel}) can be evaluated exactly~\cite{Fisher1964,Nagata1972,Note1}.
With the exchange constants as free parameters, the shift along the $x$ axis is fitted in the range $0.5~\mathrm{K} \leq T \leq 3$~K, giving $J/\kB\sim -300\pm200$~mK and $\JA/\kB\sim -9\pm4$~mK [Fig.~\ref{fig4}(b) lower solid curve].
The same parameters give a good match for the shift along the $y$ axis [Fig.~\ref{fig4}(b) upper solid curve].

As an alternative to fitting over this restricted temperature range, the data can also be fitted as a function of effective spin temperature $\Teff$ over the entire range $T_{\textrm{eff}}\leq 3$~K [lower dashed curve in Fig.~\hbox{\ref{fig4}}(b)]. This yields similar values $J/\kB=-1200\pm 500$~mK and $\JA/\kB=-10\pm 3$~mK. As before the same parameters give a good fit for the shift along $y$ [Fig.~\hbox{\ref{fig4}}(b) upper dashed curve].
Interestingly, in both cases the extracted anisotropic exchange is close to the dipole-dipole interaction $\JA/\kB = -3\mu_0 g^2\muB^2/8 \pi a^3 \kB \sim - 10$~mK estimated from the molecular spacing $a\sim 7.1$~\AA.
The deviation between fit and data presumably reflects the increasing importance of quantum correlations at low temperature and higher dimensionality of the interactions, neither of which is well captured by this one-dimensional model.
The anisotropy axis in spin resonance coincides with the long axis of the crystal (the $x$ axis) but does not appear to correspond to any preferred direction in the x-ray diffraction structure~\cite{Note1}. The temperature dependence does not simply result from a demagnetizing field, which would be weaker and would have the same sign for both orientations.

In conclusion, we have shown coupling between a microwave cavity and the molecular ensemble both in an uncorrelated and AFM correlated state~\cite{Schuster2010, Kubo2010}.
This crystal structure presumably exhibits a complex network of exchange interactions, but these circuit QED spin resonance techniques, applied in future experiments, will enable measurements of spin systems with engineered interactions, for example molecular magnets in one-dimensional chains or higher-dimensional systems with well-defined exchange pathways~\cite{GoddardPRL2012,LiuIC2016}. Magnetic resonance measurements  on these molecules offer a way to extract spin correlation functions experimentally via Eq.~(\ref{eqn:tmodel}), thereby offering a platform to test theoretical predictions for quantum correlated systems.
As a quantum memory, organic magnetic ensembles offer a high spin density, and therefore a strong ensemble coupling, with potential for chemical engineering of the spin system.

\begin{acknowledgments}
We acknowledge L.P. Kouwenhoven for use of the sputterer, C. Baines, B. Huddart, M. Worsdale, and F. Xiao for experimental assistance with muon measurements made at the Swiss Muon Source (Paul Scherrer Institut, Switzerland), S.C. Speller for discussions, and support from EPSRC (EP/J015067/1 and EP/J001821/1), Marie Curie (CIG, IEF, and IIF), the ERC (338258 ``OptoQMol"), Grant No. MAT2015-68204-R from Spanish MINECO, a Glasstone Fellowship, the Royal Society, the Royal Academy of Engineering, and Templeton World Charity Foundation. M. M. acknowledges support from the Stiftung der Deutschen Wirtschaft (sdw).
\end{acknowledgments}


\begin{thebibliography}{47}%
\makeatletter
\providecommand \@ifxundefined [1]{%
 \@ifx{#1\undefined}
}%
\providecommand \@ifnum [1]{%
 \ifnum #1\expandafter \@firstoftwo
 \else \expandafter \@secondoftwo
 \fi
}%
\providecommand \@ifx [1]{%
 \ifx #1\expandafter \@firstoftwo
 \else \expandafter \@secondoftwo
 \fi
}%
\providecommand \natexlab [1]{#1}%
\providecommand \enquote  [1]{``#1''}%
\providecommand \bibnamefont  [1]{#1}%
\providecommand \bibfnamefont [1]{#1}%
\providecommand \citenamefont [1]{#1}%
\providecommand \href@noop [0]{\@secondoftwo}%
\providecommand \href [0]{\begingroup \@sanitize@url \@href}%
\providecommand \@href[1]{\@@startlink{#1}\@@href}%
\providecommand \@@href[1]{\endgroup#1\@@endlink}%
\providecommand \@sanitize@url [0]{\catcode `\\12\catcode `\$12\catcode
  `\&12\catcode `\#12\catcode `\^12\catcode `\_12\catcode `\%12\relax}%
\providecommand \@@startlink[1]{}%
\providecommand \@@endlink[0]{}%
\providecommand \url  [0]{\begingroup\@sanitize@url \@url }%
\providecommand \@url [1]{\endgroup\@href {#1}{\urlprefix }}%
\providecommand \urlprefix  [0]{URL }%
\providecommand \Eprint [0]{\href }%
\providecommand \doibase [0]{http://dx.doi.org/}%
\providecommand \selectlanguage [0]{\@gobble}%
\providecommand \bibinfo  [0]{\@secondoftwo}%
\providecommand \bibfield  [0]{\@secondoftwo}%
\providecommand \translation [1]{[#1]}%
\providecommand \BibitemOpen [0]{}%
\providecommand \bibitemStop [0]{}%
\providecommand \bibitemNoStop [0]{.\EOS\space}%
\providecommand \EOS [0]{\spacefactor3000\relax}%
\providecommand \BibitemShut  [1]{\csname bibitem#1\endcsname}%
\let\auto@bib@innerbib\@empty
\bibitem [{\citenamefont {Wesenberg}\ \emph {et~al.}(2009)\citenamefont
  {Wesenberg}, \citenamefont {Ardavan}, \citenamefont {Briggs}, \citenamefont
  {Morton}, \citenamefont {Schoelkopf}, \citenamefont {Schuster},\ and\
  \citenamefont {M{\o}lmer}}]{Wesenberg2009}%
  \BibitemOpen
  \bibfield  {author} {\bibinfo {author} {\bibfnamefont {J.~H.}\ \bibnamefont
  {Wesenberg}}, \bibinfo {author} {\bibfnamefont {A.}~\bibnamefont {Ardavan}},
  \bibinfo {author} {\bibfnamefont {G.~A.~D.}\ \bibnamefont {Briggs}}, \bibinfo
  {author} {\bibfnamefont {J.~J.~L.}\ \bibnamefont {Morton}}, \bibinfo {author}
  {\bibfnamefont {R.~J.}\ \bibnamefont {Schoelkopf}}, \bibinfo {author}
  {\bibfnamefont {D.~I.}\ \bibnamefont {Schuster}}, \ and\ \bibinfo {author}
  {\bibfnamefont {K.}~\bibnamefont {M{\o}lmer}},\ }\href {\doibase
  10.1103/PhysRevLett.103.070502} {\bibfield  {journal} {\bibinfo  {journal}
  {Phys. Rev. Lett.}\ }\textbf {\bibinfo {volume} {103}},\ \bibinfo {pages}
  {070502} (\bibinfo {year} {2009})}\BibitemShut {NoStop}%
\bibitem [{\citenamefont {Xiang}\ \emph {et~al.}(2013)\citenamefont {Xiang},
  \citenamefont {Ashhab}, \citenamefont {You},\ and\ \citenamefont
  {Nori}}]{Xiang2013}%
  \BibitemOpen
  \bibfield  {author} {\bibinfo {author} {\bibfnamefont {Z.-L.}\ \bibnamefont
  {Xiang}}, \bibinfo {author} {\bibfnamefont {S.}~\bibnamefont {Ashhab}},
  \bibinfo {author} {\bibfnamefont {J.}~\bibnamefont {You}}, \ and\ \bibinfo
  {author} {\bibfnamefont {F.}~\bibnamefont {Nori}},\ }\href {\doibase
  10.1103/RevModPhys.85.623} {\bibfield  {journal} {\bibinfo  {journal} {Rev.
  Mod. Phys.}\ }\textbf {\bibinfo {volume} {85}},\ \bibinfo {pages} {623}
  (\bibinfo {year} {2013})}\BibitemShut {NoStop}%
\bibitem [{\citenamefont {Schuster}\ \emph
  {et~al.}(2010{\natexlab{a}})\citenamefont {Schuster}, \citenamefont
  {Fragner}, \citenamefont {Dykman}, \citenamefont {Lyon},\ and\ \citenamefont
  {Schoelkopf}}]{Schuster2010a}%
  \BibitemOpen
  \bibfield  {author} {\bibinfo {author} {\bibfnamefont {D.~I.}\ \bibnamefont
  {Schuster}}, \bibinfo {author} {\bibfnamefont {A.}~\bibnamefont {Fragner}},
  \bibinfo {author} {\bibfnamefont {M.~I.}\ \bibnamefont {Dykman}}, \bibinfo
  {author} {\bibfnamefont {S.~A.}\ \bibnamefont {Lyon}}, \ and\ \bibinfo
  {author} {\bibfnamefont {R.~J.}\ \bibnamefont {Schoelkopf}},\ }\href
  {\doibase 10.1103/PhysRevLett.105.040503} {\bibfield  {journal} {\bibinfo
  {journal} {Phys. Rev. Lett.}\ }\textbf {\bibinfo {volume} {105}},\ \bibinfo
  {pages} {040503} (\bibinfo {year} {2010}{\natexlab{a}})}\BibitemShut
  {NoStop}%
\bibitem [{\citenamefont {Kubo}\ \emph {et~al.}(2010)\citenamefont {Kubo},
  \citenamefont {Ong}, \citenamefont {Bertet}, \citenamefont {Vion},
  \citenamefont {Jacques}, \citenamefont {Zheng}, \citenamefont {Dr{\'{e}}au},
  \citenamefont {Roch}, \citenamefont {Auffeves}, \citenamefont {Jelezko},
  \citenamefont {Wrachtrup}, \citenamefont {Barthe}, \citenamefont {Bergonzo},\
  and\ \citenamefont {Esteve}}]{Kubo2010}%
  \BibitemOpen
  \bibfield  {author} {\bibinfo {author} {\bibfnamefont {Y.}~\bibnamefont
  {Kubo}}, \bibinfo {author} {\bibfnamefont {F.~R.}\ \bibnamefont {Ong}},
  \bibinfo {author} {\bibfnamefont {P.}~\bibnamefont {Bertet}}, \bibinfo
  {author} {\bibfnamefont {D.}~\bibnamefont {Vion}}, \bibinfo {author}
  {\bibfnamefont {V.}~\bibnamefont {Jacques}}, \bibinfo {author} {\bibfnamefont
  {D.}~\bibnamefont {Zheng}}, \bibinfo {author} {\bibfnamefont
  {A.}~\bibnamefont {Dr{\'{e}}au}}, \bibinfo {author} {\bibfnamefont {J.-F.}\
  \bibnamefont {Roch}}, \bibinfo {author} {\bibfnamefont {A.}~\bibnamefont
  {Auffeves}}, \bibinfo {author} {\bibfnamefont {F.}~\bibnamefont {Jelezko}},
  \bibinfo {author} {\bibfnamefont {J.}~\bibnamefont {Wrachtrup}}, \bibinfo
  {author} {\bibfnamefont {M.~F.}\ \bibnamefont {Barthe}}, \bibinfo {author}
  {\bibfnamefont {P.}~\bibnamefont {Bergonzo}}, \ and\ \bibinfo {author}
  {\bibfnamefont {D.}~\bibnamefont {Esteve}},\ }\href {\doibase
  10.1103/PhysRevLett.105.140502} {\bibfield  {journal} {\bibinfo  {journal}
  {Phys. Rev. Lett.}\ }\textbf {\bibinfo {volume} {105}},\ \bibinfo {pages}
  {140502} (\bibinfo {year} {2010})}\BibitemShut {NoStop}%
\bibitem [{\citenamefont {Ams{\"{u}}ss}\ \emph {et~al.}(2011)\citenamefont
  {Ams{\"{u}}ss}, \citenamefont {Koller}, \citenamefont {N{\"{o}}bauer},
  \citenamefont {Putz}, \citenamefont {Rotter}, \citenamefont {Sandner},
  \citenamefont {Schneider}, \citenamefont {Schramb{\"{o}}ck}, \citenamefont
  {Steinhauser}, \citenamefont {Ritsch}, \citenamefont {Schmiedmayer},\ and\
  \citenamefont {Majer}}]{Amsuss2011}%
  \BibitemOpen
  \bibfield  {author} {\bibinfo {author} {\bibfnamefont {R.}~\bibnamefont
  {Ams{\"{u}}ss}}, \bibinfo {author} {\bibfnamefont {C.}~\bibnamefont
  {Koller}}, \bibinfo {author} {\bibfnamefont {T.}~\bibnamefont
  {N{\"{o}}bauer}}, \bibinfo {author} {\bibfnamefont {S.}~\bibnamefont {Putz}},
  \bibinfo {author} {\bibfnamefont {S.}~\bibnamefont {Rotter}}, \bibinfo
  {author} {\bibfnamefont {K.}~\bibnamefont {Sandner}}, \bibinfo {author}
  {\bibfnamefont {S.}~\bibnamefont {Schneider}}, \bibinfo {author}
  {\bibfnamefont {M.}~\bibnamefont {Schramb{\"{o}}ck}}, \bibinfo {author}
  {\bibfnamefont {G.}~\bibnamefont {Steinhauser}}, \bibinfo {author}
  {\bibfnamefont {H.}~\bibnamefont {Ritsch}}, \bibinfo {author} {\bibfnamefont
  {J.}~\bibnamefont {Schmiedmayer}}, \ and\ \bibinfo {author} {\bibfnamefont
  {J.}~\bibnamefont {Majer}},\ }\href {\doibase 10.1103/PhysRevLett.107.060502}
  {\bibfield  {journal} {\bibinfo  {journal} {Phys. Rev. Lett.}\ }\textbf
  {\bibinfo {volume} {107}},\ \bibinfo {pages} {060502} (\bibinfo {year}
  {2011})}\BibitemShut {NoStop}%
\bibitem [{\citenamefont {Ranjan}\ \emph {et~al.}(2013)\citenamefont {Ranjan},
  \citenamefont {de~Lange}, \citenamefont {Schutjens}, \citenamefont
  {Debelhoir}, \citenamefont {Groen}, \citenamefont {Szombati}, \citenamefont
  {Thoen}, \citenamefont {Klapwijk}, \citenamefont {Hanson},\ and\
  \citenamefont {DiCarlo}}]{Ranjan2013}%
  \BibitemOpen
  \bibfield  {author} {\bibinfo {author} {\bibfnamefont {V.}~\bibnamefont
  {Ranjan}}, \bibinfo {author} {\bibfnamefont {G.}~\bibnamefont {de~Lange}},
  \bibinfo {author} {\bibfnamefont {R.}~\bibnamefont {Schutjens}}, \bibinfo
  {author} {\bibfnamefont {T.}~\bibnamefont {Debelhoir}}, \bibinfo {author}
  {\bibfnamefont {J.~P.}\ \bibnamefont {Groen}}, \bibinfo {author}
  {\bibfnamefont {D.}~\bibnamefont {Szombati}}, \bibinfo {author}
  {\bibfnamefont {D.~J.}\ \bibnamefont {Thoen}}, \bibinfo {author}
  {\bibfnamefont {T.~M.}\ \bibnamefont {Klapwijk}}, \bibinfo {author}
  {\bibfnamefont {R.}~\bibnamefont {Hanson}}, \ and\ \bibinfo {author}
  {\bibfnamefont {L.}~\bibnamefont {DiCarlo}},\ }\href {\doibase
  10.1103/PhysRevLett.110.067004} {\bibfield  {journal} {\bibinfo  {journal}
  {Phys. Rev. Lett.}\ }\textbf {\bibinfo {volume} {110}},\ \bibinfo {pages}
  {067004} (\bibinfo {year} {2013})}\BibitemShut {NoStop}%
\bibitem [{\citenamefont {Clauss}\ \emph {et~al.}(2013)\citenamefont {Clauss},
  \citenamefont {Bothner}, \citenamefont {Koelle}, \citenamefont {Kleiner},
  \citenamefont {Bogani}, \citenamefont {Scheffler},\ and\ \citenamefont
  {Dressel}}]{Clauss2013}%
  \BibitemOpen
  \bibfield  {author} {\bibinfo {author} {\bibfnamefont {C.}~\bibnamefont
  {Clauss}}, \bibinfo {author} {\bibfnamefont {D.}~\bibnamefont {Bothner}},
  \bibinfo {author} {\bibfnamefont {D.}~\bibnamefont {Koelle}}, \bibinfo
  {author} {\bibfnamefont {R.}~\bibnamefont {Kleiner}}, \bibinfo {author}
  {\bibfnamefont {L.}~\bibnamefont {Bogani}}, \bibinfo {author} {\bibfnamefont
  {M.}~\bibnamefont {Scheffler}}, \ and\ \bibinfo {author} {\bibfnamefont
  {M.}~\bibnamefont {Dressel}},\ }\href {\doibase 10.1063/1.4802956} {\bibfield
   {journal} {\bibinfo  {journal} {Appl. Phys. Lett.}\ }\textbf {\bibinfo
  {volume} {102}},\ \bibinfo {pages} {162601} (\bibinfo {year}
  {2013})}\BibitemShut {NoStop}%
\bibitem [{\citenamefont {Kubo}\ \emph {et~al.}(2011)\citenamefont {Kubo},
  \citenamefont {Grezes}, \citenamefont {Dewes}, \citenamefont {Umeda},
  \citenamefont {Isoya}, \citenamefont {Sumiya}, \citenamefont {Morishita},
  \citenamefont {Abe}, \citenamefont {Onoda}, \citenamefont {Ohshima},
  \citenamefont {Jacques}, \citenamefont {Dr{\'{e}}au}, \citenamefont {Roch},
  \citenamefont {Diniz}, \citenamefont {Auffeves}, \citenamefont {Vion},
  \citenamefont {Esteve},\ and\ \citenamefont {Bertet}}]{Kubo2011}%
  \BibitemOpen
  \bibfield  {author} {\bibinfo {author} {\bibfnamefont {Y.}~\bibnamefont
  {Kubo}}, \bibinfo {author} {\bibfnamefont {C.}~\bibnamefont {Grezes}},
  \bibinfo {author} {\bibfnamefont {A.}~\bibnamefont {Dewes}}, \bibinfo
  {author} {\bibfnamefont {T.}~\bibnamefont {Umeda}}, \bibinfo {author}
  {\bibfnamefont {J.}~\bibnamefont {Isoya}}, \bibinfo {author} {\bibfnamefont
  {H.}~\bibnamefont {Sumiya}}, \bibinfo {author} {\bibfnamefont
  {N.}~\bibnamefont {Morishita}}, \bibinfo {author} {\bibfnamefont
  {H.}~\bibnamefont {Abe}}, \bibinfo {author} {\bibfnamefont {S.}~\bibnamefont
  {Onoda}}, \bibinfo {author} {\bibfnamefont {T.}~\bibnamefont {Ohshima}},
  \bibinfo {author} {\bibfnamefont {V.}~\bibnamefont {Jacques}}, \bibinfo
  {author} {\bibfnamefont {A.}~\bibnamefont {Dr{\'{e}}au}}, \bibinfo {author}
  {\bibfnamefont {J.-F.}\ \bibnamefont {Roch}}, \bibinfo {author}
  {\bibfnamefont {I.}~\bibnamefont {Diniz}}, \bibinfo {author} {\bibfnamefont
  {A.}~\bibnamefont {Auffeves}}, \bibinfo {author} {\bibfnamefont
  {D.}~\bibnamefont {Vion}}, \bibinfo {author} {\bibfnamefont {D.}~\bibnamefont
  {Esteve}}, \ and\ \bibinfo {author} {\bibfnamefont {P.}~\bibnamefont
  {Bertet}},\ }\href {\doibase 10.1103/PhysRevLett.107.220501} {\bibfield
  {journal} {\bibinfo  {journal} {Phys. Rev. Lett.}\ }\textbf {\bibinfo
  {volume} {107}},\ \bibinfo {pages} {220501} (\bibinfo {year}
  {2011})}\BibitemShut {NoStop}%
\bibitem [{\citenamefont {Grezes}\ \emph {et~al.}(2015)\citenamefont {Grezes},
  \citenamefont {Julsgaard}, \citenamefont {Kubo}, \citenamefont {Ma},
  \citenamefont {Stern}, \citenamefont {Bienfait}, \citenamefont {Nakamura},
  \citenamefont {Isoya}, \citenamefont {Onoda}, \citenamefont {Ohshima},
  \citenamefont {Jacques}, \citenamefont {Vion}, \citenamefont {Esteve},
  \citenamefont {Liu}, \citenamefont {M{\o}lmer},\ and\ \citenamefont
  {Bertet}}]{Grezes2015}%
  \BibitemOpen
  \bibfield  {author} {\bibinfo {author} {\bibfnamefont {C.}~\bibnamefont
  {Grezes}}, \bibinfo {author} {\bibfnamefont {B.}~\bibnamefont {Julsgaard}},
  \bibinfo {author} {\bibfnamefont {Y.}~\bibnamefont {Kubo}}, \bibinfo {author}
  {\bibfnamefont {W.~L.}\ \bibnamefont {Ma}}, \bibinfo {author} {\bibfnamefont
  {M.}~\bibnamefont {Stern}}, \bibinfo {author} {\bibfnamefont
  {A.}~\bibnamefont {Bienfait}}, \bibinfo {author} {\bibfnamefont
  {K.}~\bibnamefont {Nakamura}}, \bibinfo {author} {\bibfnamefont
  {J.}~\bibnamefont {Isoya}}, \bibinfo {author} {\bibfnamefont
  {S.}~\bibnamefont {Onoda}}, \bibinfo {author} {\bibfnamefont
  {T.}~\bibnamefont {Ohshima}}, \bibinfo {author} {\bibfnamefont
  {V.}~\bibnamefont {Jacques}}, \bibinfo {author} {\bibfnamefont
  {D.}~\bibnamefont {Vion}}, \bibinfo {author} {\bibfnamefont {D.}~\bibnamefont
  {Esteve}}, \bibinfo {author} {\bibfnamefont {R.~B.}\ \bibnamefont {Liu}},
  \bibinfo {author} {\bibfnamefont {K.}~\bibnamefont {M{\o}lmer}}, \ and\
  \bibinfo {author} {\bibfnamefont {P.}~\bibnamefont {Bertet}},\ }\href
  {\doibase 10.1103/PhysRevA.92.020301} {\bibfield  {journal} {\bibinfo
  {journal} {Phys. Rev. A}\ }\textbf {\bibinfo {volume} {92}},\ \bibinfo
  {pages} {020301} (\bibinfo {year} {2015})}\BibitemShut {NoStop}%
\bibitem [{\citenamefont {Blum}\ \emph {et~al.}(2015)\citenamefont {Blum},
  \citenamefont {O'Brien}, \citenamefont {Lauk}, \citenamefont {Bushev},
  \citenamefont {Fleischhauer},\ and\ \citenamefont {Morigi}}]{Blum2015}%
  \BibitemOpen
  \bibfield  {author} {\bibinfo {author} {\bibfnamefont {S.}~\bibnamefont
  {Blum}}, \bibinfo {author} {\bibfnamefont {C.}~\bibnamefont {O'Brien}},
  \bibinfo {author} {\bibfnamefont {N.}~\bibnamefont {Lauk}}, \bibinfo {author}
  {\bibfnamefont {P.}~\bibnamefont {Bushev}}, \bibinfo {author} {\bibfnamefont
  {M.}~\bibnamefont {Fleischhauer}}, \ and\ \bibinfo {author} {\bibfnamefont
  {G.}~\bibnamefont {Morigi}},\ }\href {\doibase 10.1103/PhysRevA.91.033834}
  {\bibfield  {journal} {\bibinfo  {journal} {Phys. Rev. A}\ }\textbf {\bibinfo
  {volume} {91}},\ \bibinfo {pages} {033834} (\bibinfo {year}
  {2015})}\BibitemShut {NoStop}%
\bibitem [{\citenamefont {Huebl}\ \emph {et~al.}(2013)\citenamefont {Huebl},
  \citenamefont {Zollitsch}, \citenamefont {Lotze}, \citenamefont {Hocke},
  \citenamefont {Greifenstein}, \citenamefont {Marx}, \citenamefont {Gross},\
  and\ \citenamefont {Goennenwein}}]{Huebl2013}%
  \BibitemOpen
  \bibfield  {author} {\bibinfo {author} {\bibfnamefont {H.}~\bibnamefont
  {Huebl}}, \bibinfo {author} {\bibfnamefont {C.~W.}\ \bibnamefont
  {Zollitsch}}, \bibinfo {author} {\bibfnamefont {J.}~\bibnamefont {Lotze}},
  \bibinfo {author} {\bibfnamefont {F.}~\bibnamefont {Hocke}}, \bibinfo
  {author} {\bibfnamefont {M.}~\bibnamefont {Greifenstein}}, \bibinfo {author}
  {\bibfnamefont {A.}~\bibnamefont {Marx}}, \bibinfo {author} {\bibfnamefont
  {R.}~\bibnamefont {Gross}}, \ and\ \bibinfo {author} {\bibfnamefont
  {S.~T.~B.}\ \bibnamefont {Goennenwein}},\ }\href {\doibase
  10.1103/PhysRevLett.111.127003} {\bibfield  {journal} {\bibinfo  {journal}
  {Phys. Rev. Lett.}\ }\textbf {\bibinfo {volume} {111}},\ \bibinfo {pages}
  {127003} (\bibinfo {year} {2013})}\BibitemShut {NoStop}%
\bibitem [{\citenamefont {Cao}\ \emph {et~al.}(2015)\citenamefont {Cao},
  \citenamefont {Yan}, \citenamefont {Huebl}, \citenamefont {Goennenwein},\
  and\ \citenamefont {Bauer}}]{Cao2015}%
  \BibitemOpen
  \bibfield  {author} {\bibinfo {author} {\bibfnamefont {Y.}~\bibnamefont
  {Cao}}, \bibinfo {author} {\bibfnamefont {P.}~\bibnamefont {Yan}}, \bibinfo
  {author} {\bibfnamefont {H.}~\bibnamefont {Huebl}}, \bibinfo {author}
  {\bibfnamefont {S.~T.~B.}\ \bibnamefont {Goennenwein}}, \ and\ \bibinfo
  {author} {\bibfnamefont {G.~E.~W.}\ \bibnamefont {Bauer}},\ }\href {\doibase
  10.1103/PhysRevB.91.094423} {\bibfield  {journal} {\bibinfo  {journal} {Phys.
  Rev. B}\ }\textbf {\bibinfo {volume} {91}},\ \bibinfo {pages} {094423}
  (\bibinfo {year} {2015})}\BibitemShut {NoStop}%
\bibitem [{\citenamefont {Marti}\ \emph {et~al.}(2014)\citenamefont {Marti},
  \citenamefont {Fina}, \citenamefont {Frontera}, \citenamefont {Liu},
  \citenamefont {Wadley}, \citenamefont {He}, \citenamefont {Paull},
  \citenamefont {Clarkson}, \citenamefont {Kudrnovsk{\'{y}}}, \citenamefont
  {Turek}, \citenamefont {Kune{\v{s}}}, \citenamefont {Yi}, \citenamefont
  {Chu}, \citenamefont {Nelson}, \citenamefont {You}, \citenamefont {Arenholz},
  \citenamefont {Salahuddin}, \citenamefont {Fontcuberta}, \citenamefont
  {Jungwirth},\ and\ \citenamefont {Ramesh}}]{Marti2014}%
  \BibitemOpen
  \bibfield  {author} {\bibinfo {author} {\bibfnamefont {X.}~\bibnamefont
  {Marti}}, \bibinfo {author} {\bibfnamefont {I.}~\bibnamefont {Fina}},
  \bibinfo {author} {\bibfnamefont {C.}~\bibnamefont {Frontera}}, \bibinfo
  {author} {\bibfnamefont {J.}~\bibnamefont {Liu}}, \bibinfo {author}
  {\bibfnamefont {P.}~\bibnamefont {Wadley}}, \bibinfo {author} {\bibfnamefont
  {Q.}~\bibnamefont {He}}, \bibinfo {author} {\bibfnamefont {R.~J.}\
  \bibnamefont {Paull}}, \bibinfo {author} {\bibfnamefont {J.~D.}\ \bibnamefont
  {Clarkson}}, \bibinfo {author} {\bibfnamefont {J.}~\bibnamefont
  {Kudrnovsk{\'{y}}}}, \bibinfo {author} {\bibfnamefont {I.}~\bibnamefont
  {Turek}}, \bibinfo {author} {\bibfnamefont {J.}~\bibnamefont {Kune{\v{s}}}},
  \bibinfo {author} {\bibfnamefont {D.}~\bibnamefont {Yi}}, \bibinfo {author}
  {\bibfnamefont {J.-H.}\ \bibnamefont {Chu}}, \bibinfo {author} {\bibfnamefont
  {C.~T.}\ \bibnamefont {Nelson}}, \bibinfo {author} {\bibfnamefont
  {L.}~\bibnamefont {You}}, \bibinfo {author} {\bibfnamefont {E.}~\bibnamefont
  {Arenholz}}, \bibinfo {author} {\bibfnamefont {S.}~\bibnamefont
  {Salahuddin}}, \bibinfo {author} {\bibfnamefont {J.}~\bibnamefont
  {Fontcuberta}}, \bibinfo {author} {\bibfnamefont {T.}~\bibnamefont
  {Jungwirth}}, \ and\ \bibinfo {author} {\bibfnamefont {R.}~\bibnamefont
  {Ramesh}},\ }\href {http://dx.doi.org/10.1038/nmat3861
  http://10.0.4.14/nmat3861} {\bibfield  {journal} {\bibinfo  {journal} {Nat.
  Mat.}\ }\textbf {\bibinfo {volume} {13}},\ \bibinfo {pages} {367} (\bibinfo
  {year} {2014})}\BibitemShut {NoStop}%
\bibitem [{\citenamefont {Jungwirth}\ \emph {et~al.}(2016)\citenamefont
  {Jungwirth}, \citenamefont {Marti}, \citenamefont {Wadley},\ and\
  \citenamefont {Wunderlich}}]{Jungwirth2016}%
  \BibitemOpen
  \bibfield  {author} {\bibinfo {author} {\bibfnamefont {T.}~\bibnamefont
  {Jungwirth}}, \bibinfo {author} {\bibfnamefont {X.}~\bibnamefont {Marti}},
  \bibinfo {author} {\bibfnamefont {P.}~\bibnamefont {Wadley}}, \ and\ \bibinfo
  {author} {\bibfnamefont {J.}~\bibnamefont {Wunderlich}},\ }\href {\doibase
  10.1038/nnano.2016.18} {\bibfield  {journal} {\bibinfo  {journal} {Nat.
  Nanotech.}\ }\textbf {\bibinfo {volume} {11}},\ \bibinfo {pages} {231}
  (\bibinfo {year} {2016})}\BibitemShut {NoStop}%
\bibitem [{\citenamefont {Troncoso}\ \emph {et~al.}(2015)\citenamefont
  {Troncoso}, \citenamefont {Ulloa}, \citenamefont {Pesce},\ and\ \citenamefont
  {Nunez}}]{Troncoso2015}%
  \BibitemOpen
  \bibfield  {author} {\bibinfo {author} {\bibfnamefont {R.~E.}\ \bibnamefont
  {Troncoso}}, \bibinfo {author} {\bibfnamefont {C.}~\bibnamefont {Ulloa}},
  \bibinfo {author} {\bibfnamefont {F.}~\bibnamefont {Pesce}}, \ and\ \bibinfo
  {author} {\bibfnamefont {A.~S.}\ \bibnamefont {Nunez}},\ }\href {\doibase
  10.1103/PhysRevB.92.224424} {\bibfield  {journal} {\bibinfo  {journal} {Phys.
  Rev. B}\ }\textbf {\bibinfo {volume} {92}},\ \bibinfo {pages} {224424}
  (\bibinfo {year} {2015})}\BibitemShut {NoStop}%
\bibitem [{\citenamefont {Goddard}\ \emph {et~al.}(2012)\citenamefont
  {Goddard}, \citenamefont {Manson}, \citenamefont {Singleton}, \citenamefont
  {Franke}, \citenamefont {Lancaster}, \citenamefont {Steele}, \citenamefont
  {Blundell}, \citenamefont {Baines}, \citenamefont {Pratt}, \citenamefont
  {McDonald}, \citenamefont {Ayala-Valenzuela}, \citenamefont {Corbey},
  \citenamefont {Southerland}, \citenamefont {Sengupta},\ and\ \citenamefont
  {Schlueter}}]{GoddardPRL2012}%
  \BibitemOpen
  \bibfield  {author} {\bibinfo {author} {\bibfnamefont {P.~A.}\ \bibnamefont
  {Goddard}}, \bibinfo {author} {\bibfnamefont {J.~L.}\ \bibnamefont {Manson}},
  \bibinfo {author} {\bibfnamefont {J.}~\bibnamefont {Singleton}}, \bibinfo
  {author} {\bibfnamefont {I.}~\bibnamefont {Franke}}, \bibinfo {author}
  {\bibfnamefont {T.}~\bibnamefont {Lancaster}}, \bibinfo {author}
  {\bibfnamefont {A.~J.}\ \bibnamefont {Steele}}, \bibinfo {author}
  {\bibfnamefont {S.~J.}\ \bibnamefont {Blundell}}, \bibinfo {author}
  {\bibfnamefont {C.}~\bibnamefont {Baines}}, \bibinfo {author} {\bibfnamefont
  {F.~L.}\ \bibnamefont {Pratt}}, \bibinfo {author} {\bibfnamefont {R.~D.}\
  \bibnamefont {McDonald}}, \bibinfo {author} {\bibfnamefont {O.~E.}\
  \bibnamefont {Ayala-Valenzuela}}, \bibinfo {author} {\bibfnamefont {J.~F.}\
  \bibnamefont {Corbey}}, \bibinfo {author} {\bibfnamefont {H.~I.}\
  \bibnamefont {Southerland}}, \bibinfo {author} {\bibfnamefont
  {P.}~\bibnamefont {Sengupta}}, \ and\ \bibinfo {author} {\bibfnamefont
  {J.~A.}\ \bibnamefont {Schlueter}},\ }\href {\doibase
  10.1103/PhysRevLett.108.077208} {\bibfield  {journal} {\bibinfo  {journal}
  {Phys. Rev. Lett.}\ }\textbf {\bibinfo {volume} {108}},\ \bibinfo {pages}
  {077208} (\bibinfo {year} {2012})}\BibitemShut {NoStop}%
\bibitem [{\citenamefont {Liu}\ \emph {et~al.}(2016)\citenamefont {Liu},
  \citenamefont {Goddard}, \citenamefont {Singleton}, \citenamefont
  {Brambleby}, \citenamefont {Foronda}, \citenamefont {M{\"{o}}ller},
  \citenamefont {Kohama}, \citenamefont {Ghannadzadeh}, \citenamefont
  {Ardavan}, \citenamefont {Blundell}, \citenamefont {Lancaster}, \citenamefont
  {Xiao}, \citenamefont {Williams}, \citenamefont {Pratt}, \citenamefont
  {Baker}, \citenamefont {Wierschem}, \citenamefont {Lapidus}, \citenamefont
  {Stone}, \citenamefont {Stephens}, \citenamefont {Bendix}, \citenamefont
  {Woods}, \citenamefont {Carreiro}, \citenamefont {Tran}, \citenamefont
  {Villa},\ and\ \citenamefont {Manson}}]{LiuIC2016}%
  \BibitemOpen
  \bibfield  {author} {\bibinfo {author} {\bibfnamefont {J.}~\bibnamefont
  {Liu}}, \bibinfo {author} {\bibfnamefont {P.~A.}\ \bibnamefont {Goddard}},
  \bibinfo {author} {\bibfnamefont {J.}~\bibnamefont {Singleton}}, \bibinfo
  {author} {\bibfnamefont {J.}~\bibnamefont {Brambleby}}, \bibinfo {author}
  {\bibfnamefont {F.}~\bibnamefont {Foronda}}, \bibinfo {author} {\bibfnamefont
  {J.~S.}\ \bibnamefont {M{\"{o}}ller}}, \bibinfo {author} {\bibfnamefont
  {Y.}~\bibnamefont {Kohama}}, \bibinfo {author} {\bibfnamefont
  {S.}~\bibnamefont {Ghannadzadeh}}, \bibinfo {author} {\bibfnamefont
  {A.}~\bibnamefont {Ardavan}}, \bibinfo {author} {\bibfnamefont {S.~J.}\
  \bibnamefont {Blundell}}, \bibinfo {author} {\bibfnamefont {T.}~\bibnamefont
  {Lancaster}}, \bibinfo {author} {\bibfnamefont {F.}~\bibnamefont {Xiao}},
  \bibinfo {author} {\bibfnamefont {R.~C.}\ \bibnamefont {Williams}}, \bibinfo
  {author} {\bibfnamefont {F.~L.}\ \bibnamefont {Pratt}}, \bibinfo {author}
  {\bibfnamefont {P.~J.}\ \bibnamefont {Baker}}, \bibinfo {author}
  {\bibfnamefont {K.}~\bibnamefont {Wierschem}}, \bibinfo {author}
  {\bibfnamefont {S.~H.}\ \bibnamefont {Lapidus}}, \bibinfo {author}
  {\bibfnamefont {K.~H.}\ \bibnamefont {Stone}}, \bibinfo {author}
  {\bibfnamefont {P.~W.}\ \bibnamefont {Stephens}}, \bibinfo {author}
  {\bibfnamefont {J.}~\bibnamefont {Bendix}}, \bibinfo {author} {\bibfnamefont
  {T.~J.}\ \bibnamefont {Woods}}, \bibinfo {author} {\bibfnamefont {K.~E.}\
  \bibnamefont {Carreiro}}, \bibinfo {author} {\bibfnamefont {H.~E.}\
  \bibnamefont {Tran}}, \bibinfo {author} {\bibfnamefont {C.~J.}\ \bibnamefont
  {Villa}}, \ and\ \bibinfo {author} {\bibfnamefont {J.~L.}\ \bibnamefont
  {Manson}},\ }\href {\doibase 10.1021/acs.inorgchem.5b02991} {\bibfield
  {journal} {\bibinfo  {journal} {Inorganic Chemistry}\ }\textbf {\bibinfo
  {volume} {55}},\ \bibinfo {pages} {3515} (\bibinfo {year}
  {2016})}\BibitemShut {NoStop}%
\bibitem [{\citenamefont {Kahn}(1993)}]{Kahn1993a}%
  \BibitemOpen
  \bibfield  {author} {\bibinfo {author} {\bibfnamefont {O.}~\bibnamefont
  {Kahn}},\ }\href@noop {} {\emph {\bibinfo {title} {{Molecular Magnetism}}}}\
  (\bibinfo  {publisher} {VCH},\ \bibinfo {address} {New York, NY},\ \bibinfo
  {year} {1993})\BibitemShut {NoStop}%
\bibitem [{\citenamefont {Chiarelli}\ \emph {et~al.}(1993)\citenamefont
  {Chiarelli}, \citenamefont {Novak}, \citenamefont {Rassat},\ and\
  \citenamefont {Tholence}}]{Chiarelli1993}%
  \BibitemOpen
  \bibfield  {author} {\bibinfo {author} {\bibfnamefont {R.}~\bibnamefont
  {Chiarelli}}, \bibinfo {author} {\bibfnamefont {M.~A.}\ \bibnamefont
  {Novak}}, \bibinfo {author} {\bibfnamefont {A.}~\bibnamefont {Rassat}}, \
  and\ \bibinfo {author} {\bibfnamefont {J.~L.}\ \bibnamefont {Tholence}},\
  }\href {\doibase 10.1038/363147a0} {\bibfield  {journal} {\bibinfo  {journal}
  {Nature (London)}\ }\textbf {\bibinfo {volume} {363}},\ \bibinfo {pages}
  {147} (\bibinfo {year} {1993})}\BibitemShut {NoStop}%
\bibitem [{\citenamefont {Sichelschmidt}\ \emph {et~al.}(2003)\citenamefont
  {Sichelschmidt}, \citenamefont {Ivanshin}, \citenamefont {Ferstl},
  \citenamefont {Geibel},\ and\ \citenamefont {Steglich}}]{Sichelschmidt2003}%
  \BibitemOpen
  \bibfield  {author} {\bibinfo {author} {\bibfnamefont {J.}~\bibnamefont
  {Sichelschmidt}}, \bibinfo {author} {\bibfnamefont {V.~A.}\ \bibnamefont
  {Ivanshin}}, \bibinfo {author} {\bibfnamefont {J.}~\bibnamefont {Ferstl}},
  \bibinfo {author} {\bibfnamefont {C.}~\bibnamefont {Geibel}}, \ and\ \bibinfo
  {author} {\bibfnamefont {F.}~\bibnamefont {Steglich}},\ }\href {\doibase
  10.1103/PhysRevLett.91.156401} {\bibfield  {journal} {\bibinfo  {journal}
  {Phys. Rev. Lett.}\ }\textbf {\bibinfo {volume} {91}},\ \bibinfo {pages}
  {156401} (\bibinfo {year} {2003})}\BibitemShut {NoStop}%
\bibitem [{\citenamefont {Blundell}\ and\ \citenamefont
  {Pratt}(2004)}]{Blundell2004}%
  \BibitemOpen
  \bibfield  {author} {\bibinfo {author} {\bibfnamefont {S.~J.}\ \bibnamefont
  {Blundell}}\ and\ \bibinfo {author} {\bibfnamefont {F.~L.}\ \bibnamefont
  {Pratt}},\ }\href {\doibase 10.1088/0953-8984/16/24/R03} {\bibfield
  {journal} {\bibinfo  {journal} {J. Phys. Condens. Mat.}\ }\textbf {\bibinfo
  {volume} {16}},\ \bibinfo {pages} {R771} (\bibinfo {year}
  {2004})}\BibitemShut {NoStop}%
\bibitem [{\citenamefont {Wiemann}\ \emph {et~al.}(2015)\citenamefont
  {Wiemann}, \citenamefont {Simmendinger}, \citenamefont {Clauss},
  \citenamefont {Bogani}, \citenamefont {Bothner}, \citenamefont {Koelle},
  \citenamefont {Kleiner}, \citenamefont {Dressel},\ and\ \citenamefont
  {Scheffler}}]{Wiemann2015}%
  \BibitemOpen
  \bibfield  {author} {\bibinfo {author} {\bibfnamefont {Y.}~\bibnamefont
  {Wiemann}}, \bibinfo {author} {\bibfnamefont {J.}~\bibnamefont
  {Simmendinger}}, \bibinfo {author} {\bibfnamefont {C.}~\bibnamefont
  {Clauss}}, \bibinfo {author} {\bibfnamefont {L.}~\bibnamefont {Bogani}},
  \bibinfo {author} {\bibfnamefont {D.}~\bibnamefont {Bothner}}, \bibinfo
  {author} {\bibfnamefont {D.}~\bibnamefont {Koelle}}, \bibinfo {author}
  {\bibfnamefont {R.}~\bibnamefont {Kleiner}}, \bibinfo {author} {\bibfnamefont
  {M.}~\bibnamefont {Dressel}}, \ and\ \bibinfo {author} {\bibfnamefont
  {M.}~\bibnamefont {Scheffler}},\ }\href {\doibase 10.1063/1.4921231}
  {\bibfield  {journal} {\bibinfo  {journal} {Appl. Phys. Lett.}\ }\textbf
  {\bibinfo {volume} {106}},\ \bibinfo {pages} {193505} (\bibinfo {year}
  {2015})}\BibitemShut {NoStop}%
\bibitem [{Note1()}]{Note1}%
  \BibitemOpen
  \bibinfo {note} {See Supplemental Material at URL, which includes Refs.~\cite
  {Altomare1994,Parois2015,Betteridge2003,Cooper2010,Spek2003,Blundell2001,Bellido2013,Tosi2014,Jenkins2013},
  for resonator characterization, crystal characterization, susceptibility
  measurements, muon spectroscopy, data set of Crystal II and calculations of
  the temperature-dependent frequency shift, the single spin coupling and
  number of radicals in the crystal.}\BibitemShut {Stop}%
\bibitem [{\citenamefont {Prokhorov}\ and\ \citenamefont
  {Fedorov}(1963)}]{Prokhorov1963}%
  \BibitemOpen
  \bibfield  {author} {\bibinfo {author} {\bibfnamefont {A.}~\bibnamefont
  {Prokhorov}}\ and\ \bibinfo {author} {\bibfnamefont {V.}~\bibnamefont
  {Fedorov}},\ }\href
  {http://www.jetp.ac.ru/cgi-bin/e/index/e/16/6/p1489?a=list} {\bibfield
  {journal} {\bibinfo  {journal} {Soviet Phys. JETP}\ }\textbf {\bibinfo
  {volume} {16}},\ \bibinfo {pages} {1489} (\bibinfo {year}
  {1963})}\BibitemShut {NoStop}%
\bibitem [{\citenamefont {Te{\v i}tel'Baum}\ \emph {et~al.}()\citenamefont
  {Te{\v i}tel'Baum}, \citenamefont {Kharakhash'yan}, \citenamefont
  {Khlebnikov},\ and\ \citenamefont {Zenin}}]{TeitelBaum1981}%
  \BibitemOpen
  \bibfield  {author} {\bibinfo {author} {\bibfnamefont {G.~B.}\ \bibnamefont
  {Te{\v i}tel'Baum}}, \bibinfo {author} {\bibfnamefont {{\'{E}}.~G.}\
  \bibnamefont {Kharakhash'yan}}, \bibinfo {author} {\bibfnamefont {S.~Y.}\
  \bibnamefont {Khlebnikov}}, \ and\ \bibinfo {author} {\bibfnamefont {A.~G.}\
  \bibnamefont {Zenin}},\ }\href
  {http://www.jetpletters.ac.ru/ps/1535/article_23485.shtml} {\bibfield
  {journal} {\bibinfo  {journal} {JETP Letters}\ }\textbf {\bibinfo {volume}
  {34}}}\BibitemShut {NoStop}%
\bibitem [{\citenamefont {Schuster}\ \emph
  {et~al.}(2010{\natexlab{b}})\citenamefont {Schuster}, \citenamefont {Sears},
  \citenamefont {Ginossar}, \citenamefont {DiCarlo}, \citenamefont {Frunzio},
  \citenamefont {Morton}, \citenamefont {Wu}, \citenamefont {Briggs},
  \citenamefont {Buckley}, \citenamefont {Awschalom},\ and\ \citenamefont
  {Schoelkopf}}]{Schuster2010}%
  \BibitemOpen
  \bibfield  {author} {\bibinfo {author} {\bibfnamefont {D.~I.}\ \bibnamefont
  {Schuster}}, \bibinfo {author} {\bibfnamefont {A.~P.}\ \bibnamefont {Sears}},
  \bibinfo {author} {\bibfnamefont {E.}~\bibnamefont {Ginossar}}, \bibinfo
  {author} {\bibfnamefont {L.}~\bibnamefont {DiCarlo}}, \bibinfo {author}
  {\bibfnamefont {L.}~\bibnamefont {Frunzio}}, \bibinfo {author} {\bibfnamefont
  {J.~J.~L.}\ \bibnamefont {Morton}}, \bibinfo {author} {\bibfnamefont
  {H.}~\bibnamefont {Wu}}, \bibinfo {author} {\bibfnamefont {G.~A.~D.}\
  \bibnamefont {Briggs}}, \bibinfo {author} {\bibfnamefont {B.~B.}\
  \bibnamefont {Buckley}}, \bibinfo {author} {\bibfnamefont {D.~D.}\
  \bibnamefont {Awschalom}}, \ and\ \bibinfo {author} {\bibfnamefont {R.~J.}\
  \bibnamefont {Schoelkopf}},\ }\href {\doibase 10.1103/PhysRevLett.105.140501}
  {\bibfield  {journal} {\bibinfo  {journal} {Phys. Rev. Lett.}\ }\textbf
  {\bibinfo {volume} {105}},\ \bibinfo {pages} {140501} (\bibinfo {year}
  {2010}{\natexlab{b}})}\BibitemShut {NoStop}%
\bibitem [{\citenamefont {Abe}\ \emph {et~al.}(2011)\citenamefont {Abe},
  \citenamefont {Wu}, \citenamefont {Ardavan},\ and\ \citenamefont
  {Morton}}]{Abe2011}%
  \BibitemOpen
  \bibfield  {author} {\bibinfo {author} {\bibfnamefont {E.}~\bibnamefont
  {Abe}}, \bibinfo {author} {\bibfnamefont {H.}~\bibnamefont {Wu}}, \bibinfo
  {author} {\bibfnamefont {A.}~\bibnamefont {Ardavan}}, \ and\ \bibinfo
  {author} {\bibfnamefont {J.~J.~L.}\ \bibnamefont {Morton}},\ }\href {\doibase
  10.1063/1.3601930} {\bibfield  {journal} {\bibinfo  {journal} {Appl. Phys.
  Lett.}\ }\textbf {\bibinfo {volume} {98}},\ \bibinfo {pages} {251108}
  (\bibinfo {year} {2011})}\BibitemShut {NoStop}%
\bibitem [{\citenamefont {Ghirri}\ \emph {et~al.}(2015)\citenamefont {Ghirri},
  \citenamefont {Bonizzoni}, \citenamefont {Gerace}, \citenamefont {Sanna},
  \citenamefont {Cassinese},\ and\ \citenamefont {Affronte}}]{Ghirri2015b}%
  \BibitemOpen
  \bibfield  {author} {\bibinfo {author} {\bibfnamefont {A.}~\bibnamefont
  {Ghirri}}, \bibinfo {author} {\bibfnamefont {C.}~\bibnamefont {Bonizzoni}},
  \bibinfo {author} {\bibfnamefont {D.}~\bibnamefont {Gerace}}, \bibinfo
  {author} {\bibfnamefont {S.}~\bibnamefont {Sanna}}, \bibinfo {author}
  {\bibfnamefont {A.}~\bibnamefont {Cassinese}}, \ and\ \bibinfo {author}
  {\bibfnamefont {M.}~\bibnamefont {Affronte}},\ }\href {\doibase
  10.1063/1.4920930} {\bibfield  {journal} {\bibinfo  {journal} {Appl. Phys.
  Lett.}\ }\textbf {\bibinfo {volume} {106}},\ \bibinfo {pages} {184101}
  (\bibinfo {year} {2015})}\BibitemShut {NoStop}%
\bibitem [{\citenamefont {Clerk}\ \emph {et~al.}(2010)\citenamefont {Clerk},
  \citenamefont {Devoret}, \citenamefont {Girvin}, \citenamefont {Marquardt},\
  and\ \citenamefont {Schoelkopf}}]{Clerk2010}%
  \BibitemOpen
  \bibfield  {author} {\bibinfo {author} {\bibfnamefont {A.~A.}\ \bibnamefont
  {Clerk}}, \bibinfo {author} {\bibfnamefont {M.~H.}\ \bibnamefont {Devoret}},
  \bibinfo {author} {\bibfnamefont {S.~M.}\ \bibnamefont {Girvin}}, \bibinfo
  {author} {\bibfnamefont {F.}~\bibnamefont {Marquardt}}, \ and\ \bibinfo
  {author} {\bibfnamefont {R.~J.}\ \bibnamefont {Schoelkopf}},\ }\href
  {\doibase 10.1103/RevModPhys.82.1155} {\bibfield  {journal} {\bibinfo
  {journal} {Rev. Mod. Phys.}\ }\textbf {\bibinfo {volume} {82}},\ \bibinfo
  {pages} {1155} (\bibinfo {year} {2010})}\BibitemShut {NoStop}%
\bibitem [{\citenamefont {Zollitsch}\ \emph {et~al.}(2015)\citenamefont
  {Zollitsch}, \citenamefont {Mueller}, \citenamefont {Franke}, \citenamefont
  {Goennenwein}, \citenamefont {Brandt}, \citenamefont {Gross},\ and\
  \citenamefont {Huebl}}]{Zollitsch2015}%
  \BibitemOpen
  \bibfield  {author} {\bibinfo {author} {\bibfnamefont {C.~W.}\ \bibnamefont
  {Zollitsch}}, \bibinfo {author} {\bibfnamefont {K.}~\bibnamefont {Mueller}},
  \bibinfo {author} {\bibfnamefont {D.~P.}\ \bibnamefont {Franke}}, \bibinfo
  {author} {\bibfnamefont {S.~T.~B.}\ \bibnamefont {Goennenwein}}, \bibinfo
  {author} {\bibfnamefont {M.~S.}\ \bibnamefont {Brandt}}, \bibinfo {author}
  {\bibfnamefont {R.}~\bibnamefont {Gross}}, \ and\ \bibinfo {author}
  {\bibfnamefont {H.}~\bibnamefont {Huebl}},\ }\href {\doibase
  10.1063/1.4932658} {\bibfield  {journal} {\bibinfo  {journal} {Appl. Phys.
  Lett.}\ }\textbf {\bibinfo {volume} {107}},\ \bibinfo {pages} {142105}
  (\bibinfo {year} {2015})}\BibitemShut {NoStop}%
\bibitem [{\citenamefont {Katsumata}(2000)}]{Katsumata2000}%
  \BibitemOpen
  \bibfield  {author} {\bibinfo {author} {\bibfnamefont {K.}~\bibnamefont
  {Katsumata}},\ }\href {\doibase 10.1088/0953-8984/12/47/201} {\bibfield
  {journal} {\bibinfo  {journal} {J. Phys. Cond. Matter}\ }\textbf {\bibinfo
  {volume} {12}},\ \bibinfo {pages} {R589} (\bibinfo {year}
  {2000})}\BibitemShut {NoStop}%
\bibitem [{\citenamefont {Nagamiya}\ \emph {et~al.}(1955)\citenamefont
  {Nagamiya}, \citenamefont {Yosida},\ and\ \citenamefont
  {Kubo}}]{Nagamiya1955}%
  \BibitemOpen
  \bibfield  {author} {\bibinfo {author} {\bibfnamefont {T.}~\bibnamefont
  {Nagamiya}}, \bibinfo {author} {\bibfnamefont {K.}~\bibnamefont {Yosida}}, \
  and\ \bibinfo {author} {\bibfnamefont {R.}~\bibnamefont {Kubo}},\ }\href
  {\doibase 10.1080/00018735500101154} {\bibfield  {journal} {\bibinfo
  {journal} {Adv. Phys.}\ }\textbf {\bibinfo {volume} {4}},\ \bibinfo {pages}
  {1} (\bibinfo {year} {1955})}\BibitemShut {NoStop}%
\bibitem [{\citenamefont {Magari{\~{n}}o}\ \emph {et~al.}(1978)\citenamefont
  {Magari{\~{n}}o}, \citenamefont {Tuchendler},\ and\ \citenamefont
  {Renard}}]{Magarino1978}%
  \BibitemOpen
  \bibfield  {author} {\bibinfo {author} {\bibfnamefont {J.}~\bibnamefont
  {Magari{\~{n}}o}}, \bibinfo {author} {\bibfnamefont {J.}~\bibnamefont
  {Tuchendler}}, \ and\ \bibinfo {author} {\bibfnamefont {J.}~\bibnamefont
  {Renard}},\ }\href {\doibase 10.1016/0038-1098(78)90728-7} {\bibfield
  {journal} {\bibinfo  {journal} {Solid State Commun.}\ }\textbf {\bibinfo
  {volume} {26}},\ \bibinfo {pages} {721} (\bibinfo {year} {1978})}\BibitemShut
  {NoStop}%
\bibitem [{\citenamefont {Kessel}\ \emph {et~al.}(1973)\citenamefont {Kessel},
  \citenamefont {Kozyrev}, \citenamefont {Kharakhash.'yan}, \citenamefont
  {Khlebnikov},\ and\ \citenamefont {Shakirov}}]{Kessel1973}%
  \BibitemOpen
  \bibfield  {author} {\bibinfo {author} {\bibfnamefont {A.}~\bibnamefont
  {Kessel}}, \bibinfo {author} {\bibfnamefont {B.~M.}\ \bibnamefont {Kozyrev}},
  \bibinfo {author} {\bibfnamefont {E.~G.}\ \bibnamefont {Kharakhash.'yan}},
  \bibinfo {author} {\bibfnamefont {S.~Y.}\ \bibnamefont {Khlebnikov}}, \ and\
  \bibinfo {author} {\bibfnamefont {S.~Z.}\ \bibnamefont {Shakirov}},\ }\href
  {http://www.jetpletters.ac.ru/ps/1560/article_23871.shtml} {\bibfield
  {journal} {\bibinfo  {journal} {JETP Lett.}\ }\textbf {\bibinfo {volume}
  {17}},\ \bibinfo {pages} {453} (\bibinfo {year} {1973})}\BibitemShut
  {NoStop}%
\bibitem [{\citenamefont {Oshikawa}\ and\ \citenamefont
  {Affleck}(2002)}]{Oshikawa2002}%
  \BibitemOpen
  \bibfield  {author} {\bibinfo {author} {\bibfnamefont {M.}~\bibnamefont
  {Oshikawa}}\ and\ \bibinfo {author} {\bibfnamefont {I.}~\bibnamefont
  {Affleck}},\ }\href {\doibase 10.1103/PhysRevB.65.134410} {\bibfield
  {journal} {\bibinfo  {journal} {Phys. Rev. B}\ }\textbf {\bibinfo {volume}
  {65}},\ \bibinfo {pages} {134410} (\bibinfo {year} {2002})}\BibitemShut
  {NoStop}%
\bibitem [{\citenamefont {Nagata}\ and\ \citenamefont
  {Tazuke}(1972)}]{Nagata1972}%
  \BibitemOpen
  \bibfield  {author} {\bibinfo {author} {\bibfnamefont {K.}~\bibnamefont
  {Nagata}}\ and\ \bibinfo {author} {\bibfnamefont {Y.}~\bibnamefont
  {Tazuke}},\ }\href {\doibase 10.1143/JPSJ.32.337} {\bibfield  {journal}
  {\bibinfo  {journal} {J. Phys. Soc. Jpn.}\ }\textbf {\bibinfo {volume}
  {32}},\ \bibinfo {pages} {337} (\bibinfo {year} {1972})}\BibitemShut
  {NoStop}%
\bibitem [{\citenamefont {Maeda}\ and\ \citenamefont
  {Oshikawa}(2005)}]{Maeda2005}%
  \BibitemOpen
  \bibfield  {author} {\bibinfo {author} {\bibfnamefont {Y.}~\bibnamefont
  {Maeda}}\ and\ \bibinfo {author} {\bibfnamefont {M.}~\bibnamefont
  {Oshikawa}},\ }\href {\doibase 10.1143/JPSJ.74.283} {\bibfield  {journal}
  {\bibinfo  {journal} {J. Phys. Soc. Jpn.}\ }\textbf {\bibinfo {volume}
  {74}},\ \bibinfo {pages} {283} (\bibinfo {year} {2005})}\BibitemShut
  {NoStop}%
\bibitem [{\citenamefont {Fisher}(1964)}]{Fisher1964}%
  \BibitemOpen
  \bibfield  {author} {\bibinfo {author} {\bibfnamefont {M.~E.}\ \bibnamefont
  {Fisher}},\ }\href {\doibase 10.1119/1.1970340} {\bibfield  {journal}
  {\bibinfo  {journal} {Am. J. Phys.}\ }\textbf {\bibinfo {volume} {32}},\
  \bibinfo {pages} {343} (\bibinfo {year} {1964})}\BibitemShut {NoStop}%
\bibitem [{\citenamefont {Altomare}\ \emph {et~al.}(1994)\citenamefont
  {Altomare}, \citenamefont {Cascarano}, \citenamefont {Giacovazzo},
  \citenamefont {Guagliardi}, \citenamefont {Burla}, \citenamefont {Polidori},\
  and\ \citenamefont {Camalli}}]{Altomare1994}%
  \BibitemOpen
  \bibfield  {author} {\bibinfo {author} {\bibfnamefont {A.}~\bibnamefont
  {Altomare}}, \bibinfo {author} {\bibfnamefont {G.}~\bibnamefont {Cascarano}},
  \bibinfo {author} {\bibfnamefont {C.}~\bibnamefont {Giacovazzo}}, \bibinfo
  {author} {\bibfnamefont {A.}~\bibnamefont {Guagliardi}}, \bibinfo {author}
  {\bibfnamefont {M.~C.}\ \bibnamefont {Burla}}, \bibinfo {author}
  {\bibfnamefont {G.}~\bibnamefont {Polidori}}, \ and\ \bibinfo {author}
  {\bibfnamefont {M.}~\bibnamefont {Camalli}},\ }\href {\doibase
  10.1107/S0021889894000221} {\bibfield  {journal} {\bibinfo  {journal} {J.
  Appl. Crystallography}\ }\textbf {\bibinfo {volume} {27}},\ \bibinfo {pages}
  {435} (\bibinfo {year} {1994})}\BibitemShut {NoStop}%
\bibitem [{\citenamefont {Parois}\ \emph {et~al.}(2015)\citenamefont {Parois},
  \citenamefont {Cooper},\ and\ \citenamefont {Thompson}}]{Parois2015}%
  \BibitemOpen
  \bibfield  {author} {\bibinfo {author} {\bibfnamefont {P.}~\bibnamefont
  {Parois}}, \bibinfo {author} {\bibfnamefont {R.~I.}\ \bibnamefont {Cooper}},
  \ and\ \bibinfo {author} {\bibfnamefont {A.~L.}\ \bibnamefont {Thompson}},\
  }\href {\doibase 10.1186/s13065-015-0105-4} {\bibfield  {journal} {\bibinfo
  {journal} {Chem. Central J.}\ }\textbf {\bibinfo {volume} {9}},\ \bibinfo
  {pages} {30} (\bibinfo {year} {2015})}\BibitemShut {NoStop}%
\bibitem [{\citenamefont {Betteridge}\ \emph {et~al.}(2003)\citenamefont
  {Betteridge}, \citenamefont {Carruthers}, \citenamefont {Cooper},
  \citenamefont {Prout},\ and\ \citenamefont {Watkin}}]{Betteridge2003}%
  \BibitemOpen
  \bibfield  {author} {\bibinfo {author} {\bibfnamefont {P.~W.}\ \bibnamefont
  {Betteridge}}, \bibinfo {author} {\bibfnamefont {J.~R.}\ \bibnamefont
  {Carruthers}}, \bibinfo {author} {\bibfnamefont {R.~I.}\ \bibnamefont
  {Cooper}}, \bibinfo {author} {\bibfnamefont {K.}~\bibnamefont {Prout}}, \
  and\ \bibinfo {author} {\bibfnamefont {D.~J.}\ \bibnamefont {Watkin}},\
  }\href {\doibase 10.1107/S0021889803021800} {\bibfield  {journal} {\bibinfo
  {journal} {J. Appl. Crystallography}\ }\textbf {\bibinfo {volume} {36}},\
  \bibinfo {pages} {1487} (\bibinfo {year} {2003})}\BibitemShut {NoStop}%
\bibitem [{\citenamefont {Cooper}\ \emph {et~al.}(2010)\citenamefont {Cooper},
  \citenamefont {Thompson},\ and\ \citenamefont {Watkin}}]{Cooper2010}%
  \BibitemOpen
  \bibfield  {author} {\bibinfo {author} {\bibfnamefont {R.~I.}\ \bibnamefont
  {Cooper}}, \bibinfo {author} {\bibfnamefont {A.~L.}\ \bibnamefont
  {Thompson}}, \ and\ \bibinfo {author} {\bibfnamefont {D.~J.}\ \bibnamefont
  {Watkin}},\ }\href {\doibase 10.1107/S0021889810025598} {\bibfield  {journal}
  {\bibinfo  {journal} {J. Appl. Crystallography}\ }\textbf {\bibinfo {volume}
  {43}},\ \bibinfo {pages} {1100} (\bibinfo {year} {2010})}\BibitemShut
  {NoStop}%
\bibitem [{\citenamefont {Spek}(2003)}]{Spek2003}%
  \BibitemOpen
  \bibfield  {author} {\bibinfo {author} {\bibfnamefont {A.~L.}\ \bibnamefont
  {Spek}},\ }\href {\doibase 10.1107/S0021889802022112} {\bibfield  {journal}
  {\bibinfo  {journal} {J. Appl. Crystallography}\ }\textbf {\bibinfo {volume}
  {36}},\ \bibinfo {pages} {7} (\bibinfo {year} {2003})}\BibitemShut {NoStop}%
\bibitem [{\citenamefont {Blundell}()}]{Blundell2001}%
  \BibitemOpen
  \bibfield  {author} {\bibinfo {author} {\bibfnamefont {S.}~\bibnamefont
  {Blundell}},\ }\href@noop {} {}\ (\bibinfo  {publisher} {Oxford University
  Press},\ \bibinfo {address} {Oxford, England})\BibitemShut {NoStop}%
\bibitem [{\citenamefont {Bellido}\ \emph {et~al.}(2013)\citenamefont
  {Bellido}, \citenamefont {Gonz{\'{a}}lez-Monje}, \citenamefont
  {Repoll{\'{e}}s}, \citenamefont {Jenkins}, \citenamefont {Ses{\'{e}}},
  \citenamefont {Drung}, \citenamefont {Schurig}, \citenamefont {Awaga},
  \citenamefont {Luis},\ and\ \citenamefont {Ruiz-Molina}}]{Bellido2013}%
  \BibitemOpen
  \bibfield  {author} {\bibinfo {author} {\bibfnamefont {E.}~\bibnamefont
  {Bellido}}, \bibinfo {author} {\bibfnamefont {P.}~\bibnamefont
  {Gonz{\'{a}}lez-Monje}}, \bibinfo {author} {\bibfnamefont {A.}~\bibnamefont
  {Repoll{\'{e}}s}}, \bibinfo {author} {\bibfnamefont {M.}~\bibnamefont
  {Jenkins}}, \bibinfo {author} {\bibfnamefont {J.}~\bibnamefont {Ses{\'{e}}}},
  \bibinfo {author} {\bibfnamefont {D.}~\bibnamefont {Drung}}, \bibinfo
  {author} {\bibfnamefont {T.}~\bibnamefont {Schurig}}, \bibinfo {author}
  {\bibfnamefont {K.}~\bibnamefont {Awaga}}, \bibinfo {author} {\bibfnamefont
  {F.}~\bibnamefont {Luis}}, \ and\ \bibinfo {author} {\bibfnamefont
  {D.}~\bibnamefont {Ruiz-Molina}},\ }\href {\doibase 10.1039/c3nr02359a}
  {\bibfield  {journal} {\bibinfo  {journal} {Nanoscale}\ }\textbf {\bibinfo
  {volume} {5}},\ \bibinfo {pages} {12565} (\bibinfo {year}
  {2013})}\BibitemShut {NoStop}%
\bibitem [{\citenamefont {Tosi}\ \emph {et~al.}(2014)\citenamefont {Tosi},
  \citenamefont {Mohiyaddin}, \citenamefont {Huebl},\ and\ \citenamefont
  {Morello}}]{Tosi2014}%
  \BibitemOpen
  \bibfield  {author} {\bibinfo {author} {\bibfnamefont {G.}~\bibnamefont
  {Tosi}}, \bibinfo {author} {\bibfnamefont {F.~A.}\ \bibnamefont
  {Mohiyaddin}}, \bibinfo {author} {\bibfnamefont {H.}~\bibnamefont {Huebl}}, \
  and\ \bibinfo {author} {\bibfnamefont {A.}~\bibnamefont {Morello}},\ }\href
  {\doibase 10.1063/1.4893242} {\bibfield  {journal} {\bibinfo  {journal} {AIP
  Adv.}\ }\textbf {\bibinfo {volume} {4}},\ \bibinfo {pages} {087122} (\bibinfo
  {year} {2014})}\BibitemShut {NoStop}%
\bibitem [{\citenamefont {Jenkins}\ \emph {et~al.}(2013)\citenamefont
  {Jenkins}, \citenamefont {H{\"{u}}mmer}, \citenamefont {{Jos{\'{e}}
  Mart{\'{i}}nez-P{\'{e}}rez}}, \citenamefont {Garc{\'{i}}a-Ripoll},
  \citenamefont {Zueco},\ and\ \citenamefont {Luis}}]{Jenkins2013}%
  \BibitemOpen
  \bibfield  {author} {\bibinfo {author} {\bibfnamefont {M.}~\bibnamefont
  {Jenkins}}, \bibinfo {author} {\bibfnamefont {T.}~\bibnamefont
  {H{\"{u}}mmer}}, \bibinfo {author} {\bibfnamefont {M.}~\bibnamefont
  {{Jos{\'{e}} Mart{\'{i}}nez-P{\'{e}}rez}}}, \bibinfo {author} {\bibfnamefont
  {J.}~\bibnamefont {Garc{\'{i}}a-Ripoll}}, \bibinfo {author} {\bibfnamefont
  {D.}~\bibnamefont {Zueco}}, \ and\ \bibinfo {author} {\bibfnamefont
  {F.}~\bibnamefont {Luis}},\ }\href
  {http://stacks.iop.org/1367-2630/15/i=9/a=095007} {\bibfield  {journal}
  {\bibinfo  {journal} {New J. Phys.}\ }\textbf {\bibinfo {volume} {15}},\
  \bibinfo {pages} {095007} (\bibinfo {year} {2013})}\BibitemShut {NoStop}%
\end{thebibliography}
\end{document}